\documentclass[pre,showkeys,11pt]{revtex4-1}
\usepackage{hyperref}
\usepackage{amssymb}
\usepackage{amsmath}
\usepackage{mathrsfs}
\usepackage{array}
\usepackage{subfigure}
\usepackage{graphicx}
\usepackage{multirow}

\begin{document}

\title{Fluctuations, structure factor and polytetrahedra in random packings of sticky hard spheres}
\date{\today}
\author{Bl\'etry, M.}
\affiliation{ICMPE-CNRS, 2-8 rue Henri Dunant, 94320 Thiais, France}
\email[Corresponding author ]{bletry@icmpe.cnrs.fr}
\thanks{V. Russier, from ICMPE, is greatly acknowledged for his help in handling distributions and fruitful discussions, along with P. C\'en\'ed\`ese for his help in parallelizing PDF calculations, as well as A. Lemaitre, from Navier Laboratory, for fruitful discussions. }
\author{Bl\'etry, J}
\affiliation{Honorary Professor, Bahia Blanca University, Argentina}

\begin{abstract}
Sequentially-built random sphere-packings have been numerically studied in the packing fraction interval $0.329 < \gamma < 0.586$. For that purpose fast running geometrical algorithms have been designed in order to build about 300 aggregates, containing $10^6$ spheres each one, which allowed a careful study of the local fluctuations and an improved accuracy in the calculations of the pair distribution $P(r)$ and structure factors $S(Q)$ of the aggregates.

Among various parameters (Voronoi tessellation, contact coordination number distribution,...), fluctuations were quantitatively evaluated by the direct evaluation of the fluctuations of the local sphere number density, which appears to follow a power law. The FWHM of the Voronoi cells volume shows a regular variation over the whole packing fraction range.

Dirac peaks appear on the pair correlation function as the packing fraction of the aggregates decreases, indicating the growth of larger and larger polytetrahedra, which manifest in two ways on the structure factor, at low and large $Q$ values. These low PF aggregates have a composite structure  made of regular polytetrahedra embedded in a more disordered matrix. Incidentally, the irregularity index of the building tetrahedron appears as a better parameter than the packing fraction to describe various features of the aggregates structure.

\end{abstract}


\maketitle

\section{Introduction}
During the remote antiquity, corn trade was made by sacks, implicitly relying on the invariance of the grains volume to the external sack volume, i.e. on the packing fraction of the disordered packing of grains. However it was also known that the seller could win (or the buyer loose) about 10~\% if the corn were simply poured into the sack instead of being carefully shaken and densified.

Later on, the maximum value of the packing fraction of disordered packings of (sticky) hard spheres--or random close packing--has been experimentally measured between 0.636 and 0.64 \cite{S60,B62,B83}. However, this value lacks any mathematical demonstration, by contrast with the case of periodic or crystalline arrangement of spheres for which it was recently shown that the maximum packing fraction is $\pi/\sqrt{18}\approx 0.74$ (Kepler conjecture demonstrated by TC Hales \cite{H02} and still being verified by several mathematician teams).

On the other hand, the onset of electronic computers about 50 years ago allowed this problem to be numerically tackled. 
Schematically, two broad families of random aggregates building families exist. The most widely used nowadays--the literature is too abundant to be exhaustively mentioned here--is the family of "dynamic" methods, for which all spheres in the aggregate are included since the beginning, and the system evolves towards equilibrium either by solving equation of motion (eg molecular dynamics \cite{AW59}, Lubachevsky-Stillinger algorithm \cite{LS90}) or on the basis of purely geometrical constraints (eg Jodrey-Tory algorithm \cite{JT85}). 
The second family, that of static--or sequential--methods, is based on the progressive insertion of spheres in the aggregate, tangentially to three already inserted spheres. In this case, the sphere is immediately assigned its definitive position and various strategies exist to build random systems in this way (\cite{JPM92,TS04}). Such approaches have been proven able to describe the structure of pure or binary liquids and amorphous metals and alloys \cite{LSSB82} and are of interest to describe penetration \cite{MJ90}, segregation effect \cite{JM90}, growth of tumor \cite{E61}...

It turns out that aggregates produced by either family of building-method differ at least in one perspective: the average contact coordination number (CCN) varies roughly between 4 and 7 (see eg \cite{BM60,S62,ASS06,WSJM11}) from the RLP to the RCP packing fraction for dynamic systems, whereas sequential methods produce aggregates with an average CCN of 6, whatever the packing fraction \cite{B72,JPM92}. 
Hence, it seems that sequential methods give access to a family of random aggregates that significantly differ from the ones obtained by dynamic methods. 

If various approaches have allowed a systematic study of dynamically built random aggregates by controlling some rate parameters to vary progressively the packing fraction of such systems (see, for example, \cite{TS10}), to the best of our knowledge, such study does not exist in the case of sequentially built random aggregates. 
In their investigation of packings built with sequential models, Jullien et al \cite{JPM92} were able to produce 5 types of aggregates whose packing fraction varied from 0.5447 to 0.6053 by changing the building procedure (Bennet method, ballistic, anti-Bennet, stable Eden, and Eden methods).
The aim of the present study is to analyse several families of sequentially built random aggregates of large number ($10^6$) spheres, whose packing fraction can be controlled by varying a continuous parameter. These geometrical results will also be of interest to interpret some structural signatures in a more general perspective.

\section{Methods}
\subsection{Building the aggregate}
\subsubsection{Sphere positioning algorithms}
\label{sec:alg}
Each spherical "aggregate" or "cluster" with radius $R$ is built by adding spheres (with diameter $d=2$ or radius $r_s = 1$ in arbitrary unit of length) one by one to the growing aggregate. In order to determine the three coordinates of the new sphere center $P$, each sphere is brought tangentially to three already positioned spheres forming a "triplet".

This triplet is formed from an "origin" sphere (center) $O$ and a pair of spheres (centers) $A$ and $B$ belonging to the neighbourhood of $O$.
This origin $O$ is randomly chosen among all the possible origins belonging to the cluster (i.e. spheres with less than 12 contacting neighbours, 12 being the maximum possible number of spheres that can contact a given sphere).

The pair of sphere centers $A$, $B$ that will form, together with $O$, the positioning triangle $OAB$ of $P$ must be chosen in the neighbourhood of $O$. More accurately, $A$ and $B$ must be contained within a cube centered on $O$ whose edge length can vary continuously from $5$ to $9$ $r_s$.  
Furthermore, within this cube, the pair of spheres $A$ and $B$ can be chosen in two ways : 

1) randomly (algorithm RAND)

2) or by choosing the triplet that maximises the sum of the three distances between $O$, $A$ and $B$ ($OA + OB + AB$)
i.e. the triplet that forms the largest hole (after ordering all the possible triplets in the cube) (algorithm MAX). 

Finally, the triplet selection procedure can be repeated for each addition of a new sphere, or it is possible to try to insert up to 9 new spheres around the same origin $O$ by selecting new pairs of spheres $A$ and $B$ in the same local cube, thus giving rise to sub-algorithms RAND-1 to RAND-9 and MAX-1 to MAX-9.

The densest structures are constructed in approximately 500~s, and only 50~s for the least dense ones, using intel i7 cpu.

\subsubsection{Aggregate radius and packing fraction}

\begin{figure}[htbp]
\begin{center}
\includegraphics[width=0.8\textwidth]{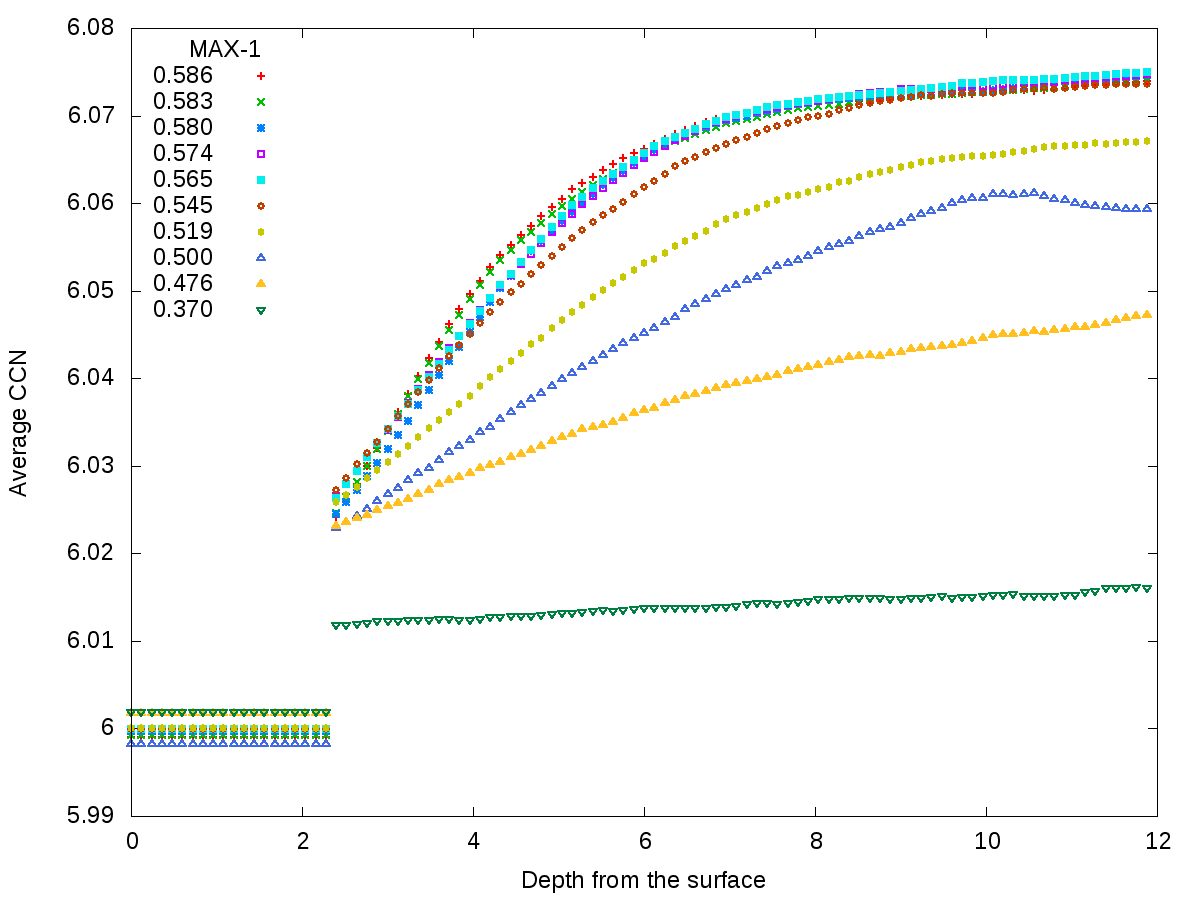}
\caption{Variation of $\bar\eta$ as a function of the thickness of the layer removed from the aggregate in $r_s$ unit.\label{cooRfluc}}
\end{center}
\end{figure}

The distribution of the contact coordination numbers ($\eta$) of all the spheres within an aggregate is directly obtained from the building process of the aggregate, and the average of $\eta$, $\bar\eta$,  over all spheres is the average contact coordination number (CCN). However, sphere centers lying near the aggregate surface have a lower CCN than the bulk ones. Figure \ref{cooRfluc} shows that this effect slightly decreases $\bar\eta$, by about 1~\%, and does not extend beyond a depth of about $3d$. In order to get rid of this surface effect, that layer is removed when calculating the average CCN, which also improves the aggregate sphericity. As a matter of fact, this sphericity will be shown to be critical in the calculations presented hereafter.

A first approximation of the aggregate radius is calculated from the average distance from the origin of all the sphere centers $R_i$ in the aggregate by relation:
\begin{equation}
R = \frac{4}{3}\frac{1}{N}\sum R_i
\end{equation}
valid for large sphere numbers.

By using these algorithms, more than 300 aggregates were built, each one containing $N=10^6$ spheres, with cluster radii $R$ in the interval $58d$ to $100d$, and packing fraction $\gamma = Nd^3/8R^3$ ranging from $0.329$ up to $0.586$. 

\begin{figure}[htbp]
\begin{center}
\includegraphics[width=0.8\textwidth]{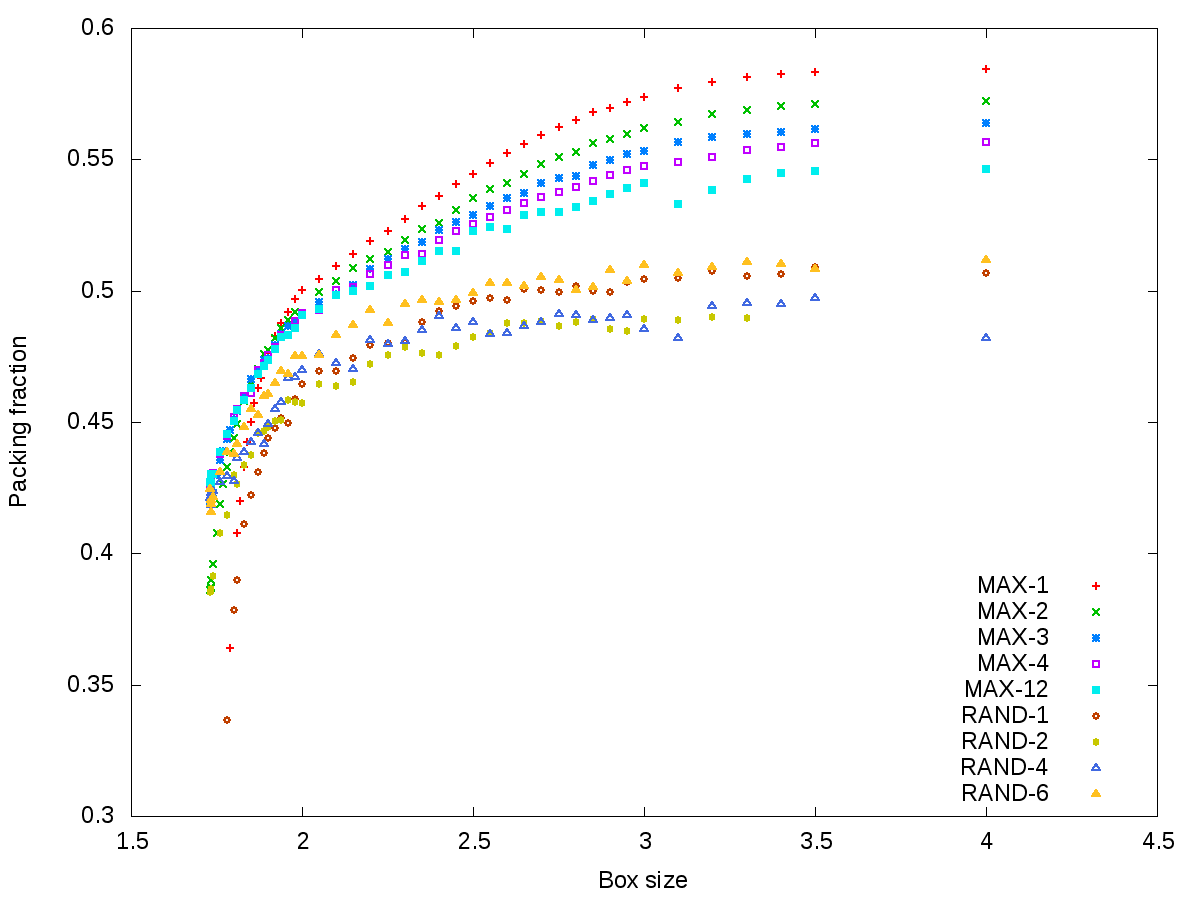}
\caption{Packing fraction as a function of the half box-edge length for algorithms MAX and RAND with the insertion of 1 up to 12 spheres around an origin.\label{fig:cuComp}}
\end{center}
\end{figure}

The packing fraction is a key parameter to classify aggregates although it does not determine unequivocally their structure, which depends on many short and medium range order parameters. It turns out that for a given packing algorithm, the packing fraction increases when the box size increases. Moreover, MAX algorithms for a given box size generate denser aggregates than RAND algorithms, as they optimize positions up to the second neighbours. Finally, for sub-algorithms MAX-1 to MAX-9, increasing the insertion number around a given origin decreases the overall packing fraction (cf. fig \ref{fig:cuComp}).

\subsubsection{Irregularity index of the sphere positioning tetrahedra}
Each added sphere forms a more or less regular tetrahedron 
$POAB$ with its positioning triplet $O$,$A$,$B$. This tetrahedron owns:
\begin{itemize}
\item Three edges $PO$, $PA$, $PB$ with length $l = d$ corresponding to the three contacts of $P$ with $O$, $A$ and $B$. These three edges contribute to the contacting neighbours Dirac peak at $\delta(r-d)$ in the pair distribution function $P(r)$ (see section \ref{sec:pdr}).
\item Three edges with length $2d \geq l > d$ corresponding to the base triangle $OAB$. These three edges contribute to the "nearby first neighbours" in $P(r)$ (see section \ref{sec:fn}).
\end{itemize}
The irregularity of the positioning tetrahedron $POAB$ can therefore be characterized by the "irregularity index":
\begin{equation}
\kappa = \frac{3d^2 + OA^2 + OB^2 + AB^2}{6d^2}
\label{perfind}
\end{equation}
$\kappa=1$ for a regular tetrahedron with 6 edges of length $d$, and $\kappa>1$ for an irregular tetrahedron. 
$\kappa$ reaches a maximum value of 2 for the "planar" tetrahedron formed by inserting the new sphere in a equilateral triplet with edge $d\sqrt{3}$, which is also the maximum hole size that can be filled by a sphere.
The average tetrahedron irregularity index, $\bar\kappa$, of an aggregate is calculated by averaging $\kappa$ over all the sphere positioning tetrahedra.

\subsubsection{Calculating the number density fluctuations}
The fluctuations of the sphere number density within the aggregates were directly derived from the positions of the spheres centers. In practice, a large cube of edge $50d$ centered on the aggregate origin $(0,0,0)$ is subdivided into $1000$ subcubes of edge $5d$, each of them containing $n$ spheres centers with an average value $\bar{n}$ and a mean square deviation $\overline{(n-\bar n)^2}/\bar{n}$ (both averaged over these $1000$ subcubes), characterizing the number density fluctuations in the aggregate. As an order of magnitude $\bar{n}\approx125\times 6\gamma/\pi\approx 100$ sphere centers are contained in each subcube and the error on the mean square standard deviation is about 10~\%.

\subsection{Voronoi tessellation}
The Voronoi tessellation of the aggregates were built thanks to the Voro++ library \cite{R09}, which also provided the number of faces and the volumes of the Voronoi cells associated with each sphere.

\subsection{Pair distribution function}
\label{sec:pdr}
The pair distribution (or correlation) functions (PDF) of the spheres belonging to a given aggregate, i.e. the probability of finding a sphere center at a distance lying between $r$ and $r+ \Delta r$ from another sphere center, normalized to $1$ as $r$ goes to infinity, is given by:
\begin{equation}
P(r) = \frac{\Delta \mathscr{P}}{\Delta r}\rho^{-2}\frac{1}{\mathscr{S}} 
\label{pdrnum}
\end{equation}
where $\Delta \mathscr{P}(r)$ is the number of sphere center pairs lying between $r$ and $r+\Delta r$, $\rho = N/V$ is the aggregate number density, $V = 4\pi R^3/3$ is the aggregate volume and $\mathscr{S} = \frac{\pi^2}{6}r^2(2R-r)^2(4R+r)$ is the spherical shape factor of the aggregate \cite{B90}. The $\Delta r$ step used in the present calculations was taken as $\sigma = 0.005 d = 0.01r_s$. The precision of the results for $P(r)$ are of the order of $2.10^{-4}$, according to the relation derived in \cite{B79}:
\begin{equation}
\frac{\Delta P}{P} = \left( \frac{16}{15} \frac{d}{\sigma}\right)^{1/2} \gamma^{-1/6} N^{-5/6}
\label{pdrnum2}
\end{equation}

\subsubsection{Renormalization of P(r)}
\label{sec:renormpr}
The pair distribution function $P(r)$ calculated by relation \ref{pdrnum} should go to $1$ as $r$ goes to infinity. However, in spite of the suppression of the surface layer, the external shapes of the aggregates are not perfectly spherical and an effective aggregate radius $R' = R(1+\varepsilon)$ must be calculated in order to properly normalize $P'(r)$ to $1$ as $r \rightarrow \infty$. If $P'(r) \approx 1$, then: 
\begin{equation}
\varepsilon(r)=\frac{1-P(r)}{P(r)} \frac{16R^3 - 12R^2r +r^3}{48R^3 - 12R^2r + 6r^3}
\label{renormpdr}
\end{equation}
$\varepsilon(r)$ value has been averaged in the interval $10d$ to $50d$ where $P(r)$ oscillations around $1$ almost vanish and $P(r)\approx 1$.
It turns out that this necessary $R$ correction does not exceed $10^{-3}$.

\subsubsection{Dirac peak of contacting neighbours}
\begin{figure}[htbp]
\begin{center}
\includegraphics[width=0.8\textwidth]{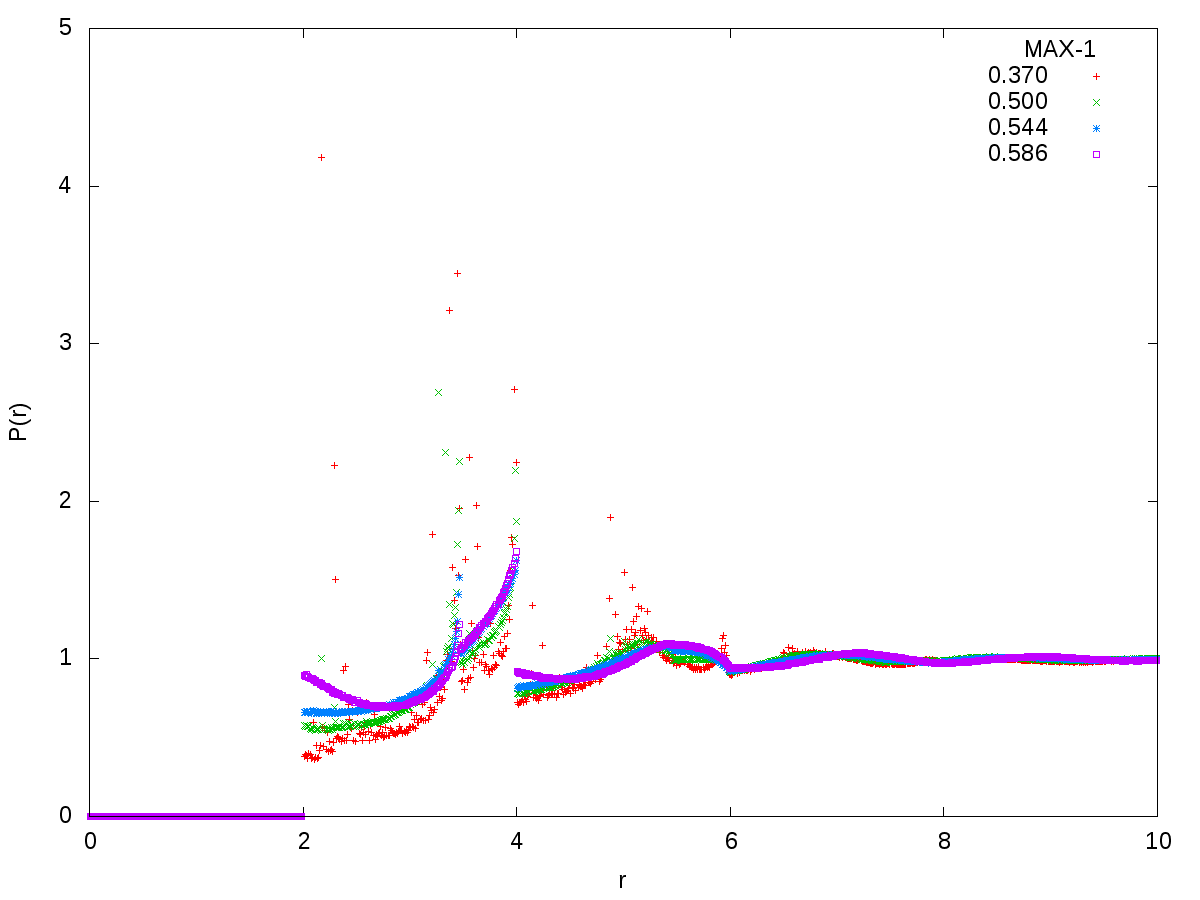}
\caption{Typical renormalized $P(r)$ curves obtained for various packing fractions for the MAX-1 algorithm. The Dirac peak at $r = d$, corresponding to contacting neighbours, falls out of range and has been skipped. Its evolution is representative of all other families of algorithms.\label{pdrtyp}}
\end{center}
\end{figure}

Typical pair distribution functions are presented in figure \ref{pdrtyp} for different packing fractions in the case of the MAX-1 algorithm. 
$P(r)$ functions are obviously null for $r< d$ and exhibit a first Dirac peak at $r = d$ whose intensity is proportional to the average number of spheres contacting a given sphere (or contact coordination number), $\bar\eta$, according to the analytical relation \cite{B90}:
\begin{equation}
P(r) = \frac{\bar\eta}{4\pi\rho d^2}\delta(r-d)
\label{deltapvan}
\end{equation}
Numerical $P(d)$, given by $P(d) = \bar\eta d/24\gamma\sigma$, falls out of range of figure \ref{pdrtyp} and will be skipped in the following figures. 
However, $\bar\eta$ deduced from $P(d)$ may slightly depart from the one directly obtained from the sphere positioning algorithm by averaging the contact coordination numbers of all spheres.

\subsection{Structure factor}
\label{sec:cordif}
Sphere aggregates are used for the simulation of the structure of disordered materials (liquid or amorphous) which are experimentally studied by diffraction experiments. 
The intensity diffracted by a sphere aggregate in the direction of the scattering vector $\mathbf Q$ is given by:
\begin{equation}
I(\mathbf{Q}) = \frac{1}{N} \sum_{i,j}\exp(i\mathbf{Q}.(\mathbf{R}_{i} - \mathbf{R}_{j}))
\label{eq:scinInt}
\end{equation}
where the sums extend over the centers positions $\mathbf{R}_{i}$ and $\mathbf{R}_{j}$ of the $N$ spheres. For a disordered system, $I(\mathbf{Q})$ should only depend on $\mathbf{Q}$ modulus but not on its orientation. In that case, Guinier \cite{G94} has shown that the scattered intensity can be (approximately) split in two parts according to relation:
\begin{equation}
I(\mathbf{Q}) \approx N\left| \Phi(Q) \right|^2 + S(Q)
\label{eq:scinap}
\end{equation}
where
\begin{equation}
\Phi(Q) = 3\frac{\sin(QR) - QR\cos(QR)}{(QR)^3}
\label{eq:scinap2}
\end{equation}
is the small angle scattering term related to the external spherical shape of the aggregate, and:
\begin{equation}
S(Q) = 1 + \frac{6\gamma}{\pi d^3} \int_0^\infty \frac{\sin(Qr)}{Qr} \left[ P(r)-1 \right] 4\pi r^2 dr
\label{eq:scinap3}
\end{equation}
is the structure factor associated with the inner structure of the aggregate.

\paragraph{Small $Q$ behaviour} Of particular interest, is the small $Q$ behaviour of $S(Q)$ for $Qd << 1$, which, according to Ornstein and Zernike \cite{OF14}, is related to the number density fluctuations by:
\label{sec:sinap4}
\begin{equation}
S(0) = \frac{\overline{(n-\bar n)^2}}{\bar{n}}
\label{eq:scinap4}
\end{equation}

\paragraph{Asymptotic behaviour} Another range of interest is the asymptotic behaviour of $S(Q)$ for $Qd >> 1$ . If the Dirac peak of the contacting neighbours in $P(r)$ is the only $\delta$ peak in $P(r)$, it determines the asymptotic behaviour of $S(Q)$ according to \cite{B77}:
\begin{equation}
S(Q) = 1+\bar\eta\frac{\sin(Qd)}{Qd}
\label{eq:scinap5}
\end{equation}

In order to finally extract $S(Q)$, we may either calculate it from relation \ref{eq:scinInt} ("interference method") or from relation \ref{eq:scinap3}, while limiting its integration range to the available $r$ interval $[0, 2R]$ ("Pair distribution or $P(r)$ method"). Both methods are detailed hereafter.

\subsubsection{Interference method}
\paragraph{Orientational effect} Although the disordered aggregates contain $10^6$ spheres, they are not fully isotropic and the scattered intensity depends on $\mathbf Q$ orientation. In order to correct this effect, relation \ref{eq:scinInt} was averaged over 728 random orientations for each $Q$ value. This process is quite time consuming and has only been used in the interval $ 0 < Q < Q_1$ (where $Q_1$ is the position of the first peak of $S(Q)$) aiming at the accurate calculation of the structure factor for small $Q$ values.

\paragraph{Small angle scattering correction} In order to extract $S(Q)$ from relations \ref{eq:scinInt} and \ref{eq:scinap}, $I(Q)$ has to be corrected for the small angle term \ref{eq:scinap2} whose amplitude is very large in the $0 < QR < 1$ interval since it is proportional to the total number of spheres in the aggregate.

$S(Q)$ could be accurately determined down to $Qd \approx 0.8$ and, in the interval $0 < Qd < 0.8$, was extrapolated to the value $S(0)$ calculated from the number density fluctuations (relation \ref{eq:scinap4}).

\subsubsection{P(r) method}
\label{sssec:pr}
\paragraph{Small angle scattering correction} $S(Q)$ cannot be calculated exactly from $P(r)$ since the aggregate has a finite size and the upper integration bound in equation \ref{eq:scinap3} is limited to $2R$. However this limitation can be approximately compensated by the small angle scattering correction given by equation \ref{eq:scinap2}.

\paragraph{Renormalization of $P(r)$} Finally, a new drawback appears. As already seen in section \ref{sec:renormpr}, the aggregate radius can be determined with an accuracy of $10^{-3}$ by a correction based on $P(r)$ normalization to 1 at large $r$. However, this is still insufficient. Indeed, if $P(r)$ goes to $1+\varepsilon$, where $\varepsilon$ is very small with respect to 1, as $r \rightarrow \infty$, an additional term of:
\begin{equation}
4\pi \rho \varepsilon \frac{\sin(2QR) - 2QR\cos(2QR)}{Q^3}
\end{equation}
is added to $S(Q)$ and produces parasitic oscillations with (pseudo)period $\pi/R$ which disturb the calculation of $S(Q)$ in the low $Q$ region. These oscillations could be suppressed by adjusting $R$ thanks to a minimization method with an accuracy of about $10^{-4}$.

\vspace{5mm}

\begin{figure}[htbp]
\begin{center}
\includegraphics[width=0.8\textwidth]{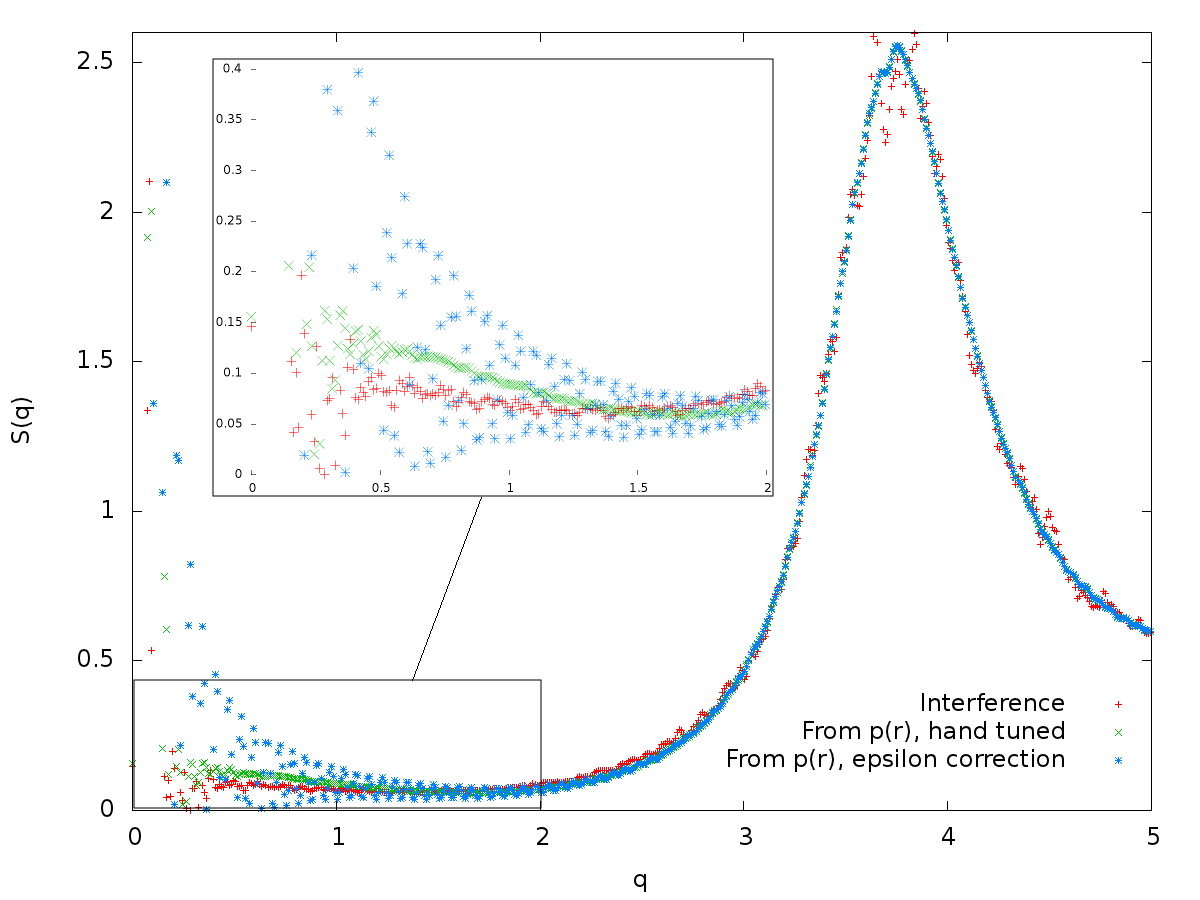}
\caption{Small $Q$ behaviour of the structure factor. Comparison of the structure factors determined by relation \ref{eq:scinInt} (labelled "interference") or from $P(r)$ (equation \ref{eq:scinap3}). 
The most noisy curve is obtained by renormalizing $R$ and $P(r)$ according to relation \ref{renormpdr} with a precision of $10^{-3}$ while the optimized curve is obtained by renormalizing $R$ and $P(r)$ with a precision of $10^{-4}$. The value of $S(0)$ is deduced from the sphere number density fluctuations (relation \ref{eq:scinap4}). \label{int_vs_pdr}}
\end{center}
\end{figure}

For both interference and $P(r)$ methods, the $\Delta Q$ step chosen was $0.02/d = 0.01$~a.u.$^{-1}$.
In practice, both methods agree satisfactorily. Figure \ref{int_vs_pdr} presents typical structure factor in the low $Q$ regime calculated by both methods, as well as the effect of the low $Q$ correction.

\section{Results and discussion}

\subsection{Packing fractions}
By using the different sphere packing algorithms, the packing fraction could be varied between 0.329 and 0.586. As a matter of fact, the RAND family of algorithms is bounded by an upper limit of $\gamma\approx 0.51$ while a maximum value of 0.586 could be reached by the MAX family of algorithms. However, the maximum RCP value of 0.636 could not be reached.
This agrees with the observations on bead packs by Aste et al. \cite{ASSS04}, who conclude that beyond $\gamma \approx 0.6$ the densification can occur by "collective and correlated readjustments of larger sets of spheres", which is precisely what static methods cannot do.
One can also notice that To et al. \cite{TS04}, using similar geometrical algorithms, did not go beyond $\gamma = 0.603$ and Jullien et al \cite{JPM92}, beyond $\gamma = 0.6053$.
Finally, it may be noticed that Ichikawa \cite{I75}, using a modified Bennett method, has reached the value of 0.627 by early computer modeling. However, this last result might suffer a large uncertainty due to the small number of spheres in the aggregates (1700), with about 50~\% of the spheres on the surface of the aggregate.

\subsection{Average irregularity index of the sphere positioning tetrahedra}
\begin{figure}[htbp]
\begin{center}
\includegraphics[width=0.8\textwidth]{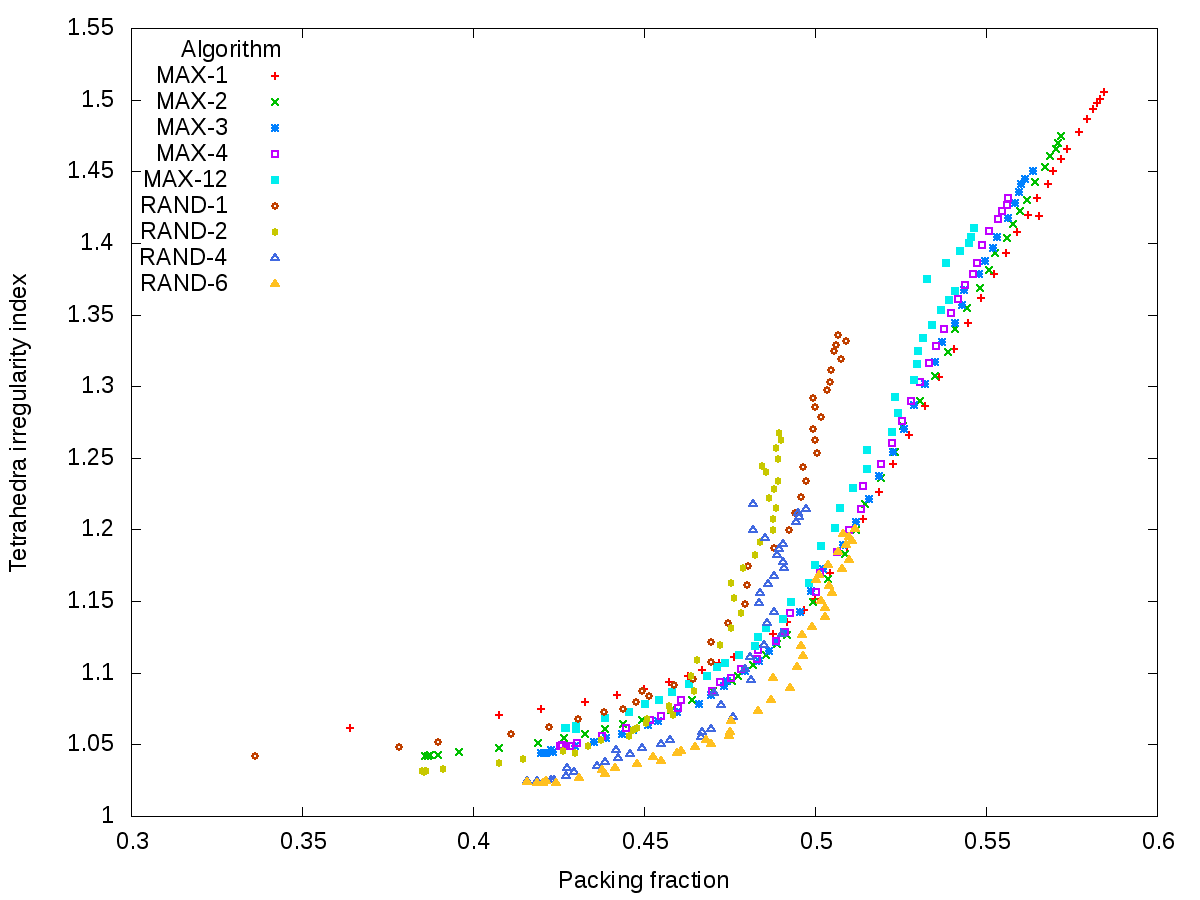}
\caption{Variations of the average tetrahedron irregularity index $\bar\kappa$ with packing fraction
\label{iperf_v_compa}}
\end{center}
\end{figure}

The tetrahedron irregularity index averaged over all sphere positioning tetrahedra, $\bar\kappa$, generally increases with packing fraction (figure \ref{iperf_v_compa}). This result indicates that the distortion of the positioning tetrahedra is a major factor controlling the packing fraction.
On the other hand, for all packing algorithms, there is a change in the almost linear growth rate of $\gamma$ as a function of $\bar\kappa$ around $\bar\kappa \approx 1.1$ and for $\gamma$ lying in the interval $0.46<\gamma<0.5$. 

\subsection{Fluctuations}
\subsubsection{Number density fluctuations}
\begin{figure}[htbp]
\begin{center}
\includegraphics[width=0.8\textwidth]{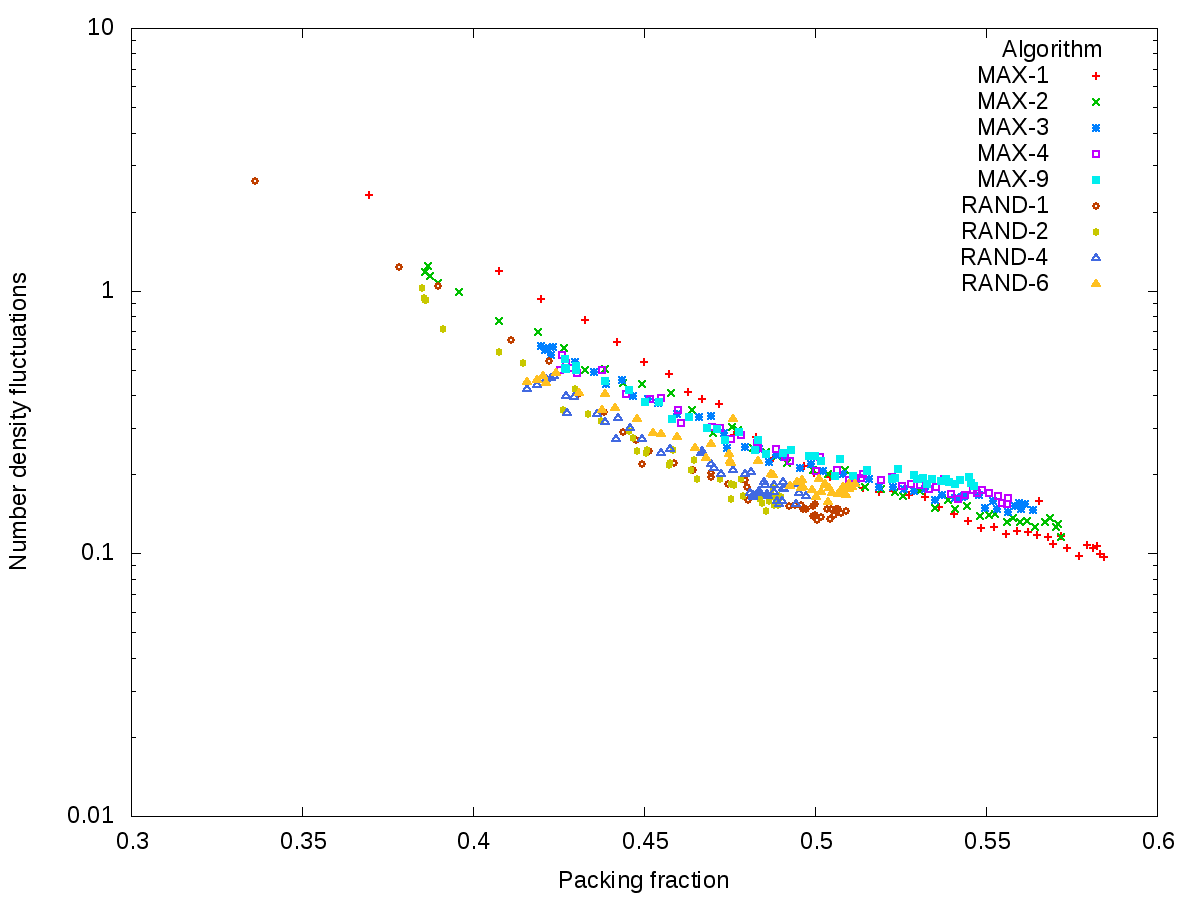}
\caption{Variations of the mean square deviation $S(0)$ of the sphere number density with packing fraction. \label{fluccompa}}
\end{center}
\end{figure}

The mean square fluctuation of the sphere number density given by equation \ref{eq:scinap4} decreases with the packing fraction and follows approximately a power law with exponent -3 (see figure \ref{fluccompa}), independently of the algorithm used. Extrapolating this last result suggests that at RCP the value of these fluctuations should be below $10^{-2}$ or $10^{-3}$.

\subsection{Contact coordination number distributions}
\begin{figure}[htbp]
\begin{center}
\includegraphics[width=0.8\textwidth]{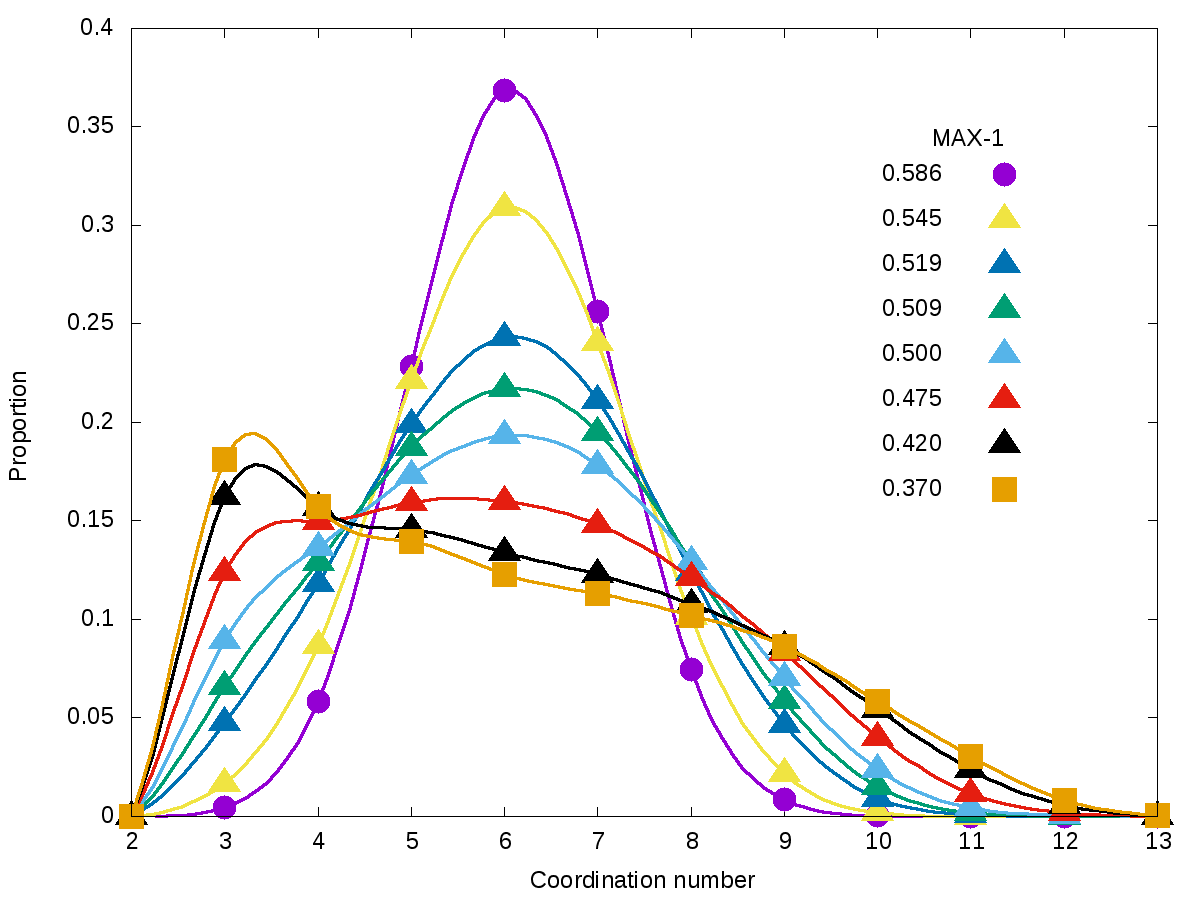}
\caption{Typical distributions of the contact coordination number for RAND-1. All other algorithms display the same evolution of the distribution but the apparition of the low coordination number peak occurs at different packing fraction. Lines are guides for the eye.\label{dist}}
\end{center}
\end{figure}

Typical distributions of the CCN are displayed in figure \ref{dist} for various packing fractions. This distribution evolves with decreasing packing fraction, from a single peak distribution to a double peaked one. This second peak appears at $\eta = 3$. The transition between these two regimes intervenes at a packing fraction value which depends on the algorithm and lies in the interval $\gamma\approx0.48$ for RAND-1 to $\gamma\approx0.52$ for MAX-1.

\begin{figure}[htbp]
\begin{center}
\includegraphics[width=0.8\textwidth]{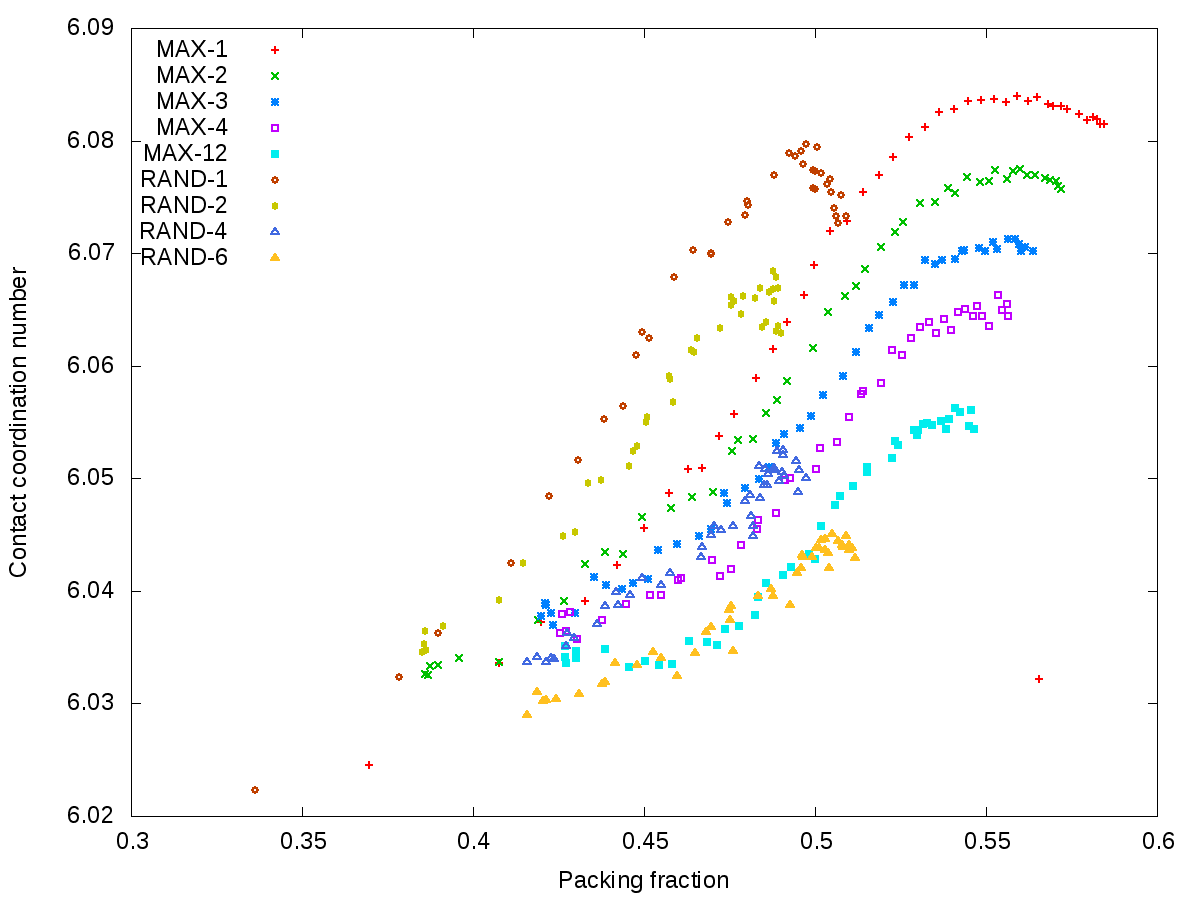}
\caption{Average CCN as a function of packing fraction.\label{avgeCCN}}
\end{center}
\end{figure}

On the other hand, the average CCN varies slightly and increases linearly with packing fraction (figure \ref{avgeCCN}), whatever the algorithm, and reaches $6.07$ for $\gamma = 0.586$. The relative constancy of $\bar\eta$ occurs for an extended packing fraction and is due to a compensation between low (3-4) and high coordination number (9 and above) in the distribution curve, whose proportion increases when the packing fraction decreases.

\subsection{Voronoi tessellation}
\label{sec:voro}
Voronoi tessellations of 2D and 3D random aggregates have been widely studied, both from an experimental and theoretical perspective (e.g. \cite{F70,WKW86,PBC04,TS04,SWJM10}).

\begin{figure}[htbp]
\begin{center}
\includegraphics[width=0.8\textwidth]{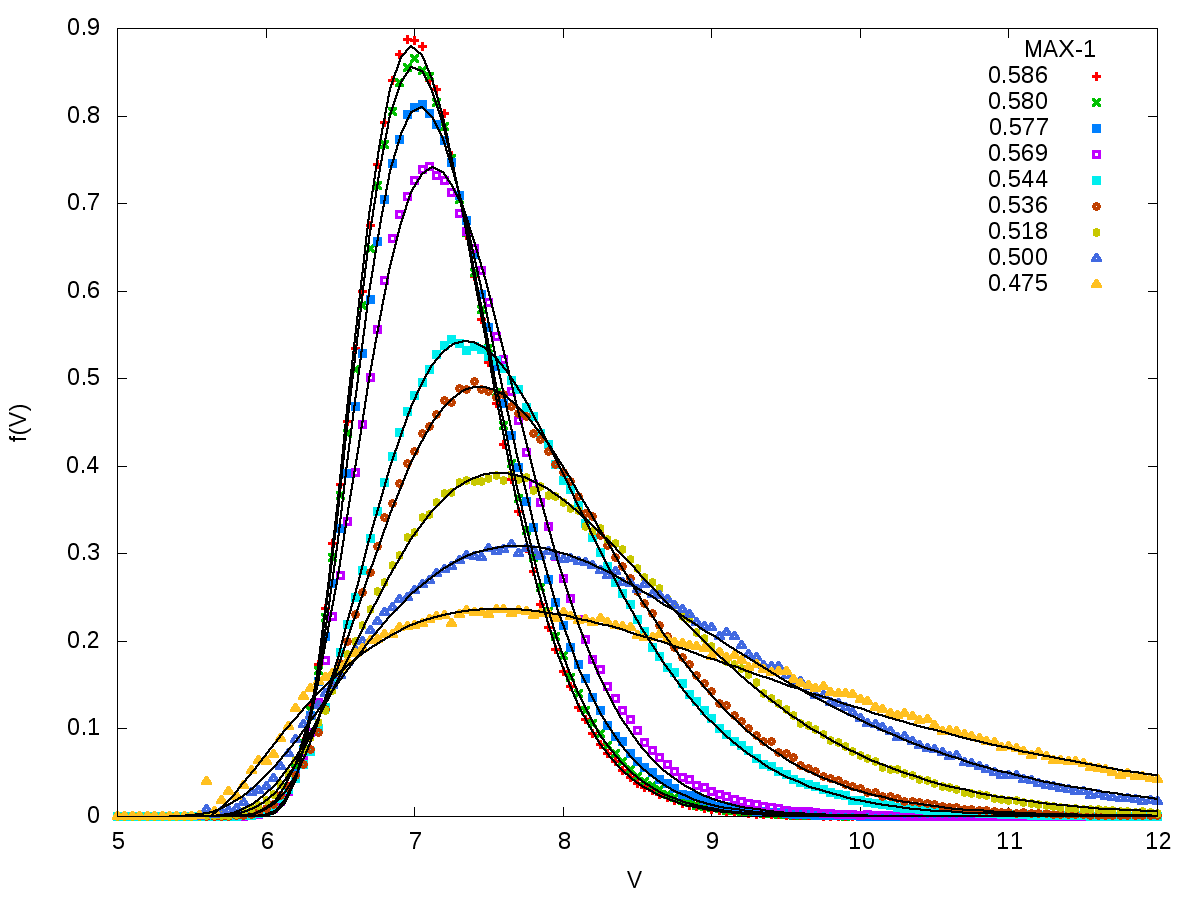}
\caption{Proportion of Voronoi cells volume $f(V)$ for aggregates of various packing fractions. V is in $r_s^3$. Black lines are fit to k-Gamma distributions (see equation \ref{eq:kgamma}).\label{fig:distvo}}
\end{center}
\end{figure}

The distributions of the volumes of the Voronoi cells for various packing fractions are presented in figure \ref{fig:distvo}. 

\begin{figure}[htbp]
\begin{center}
\includegraphics[width=0.8\textwidth]{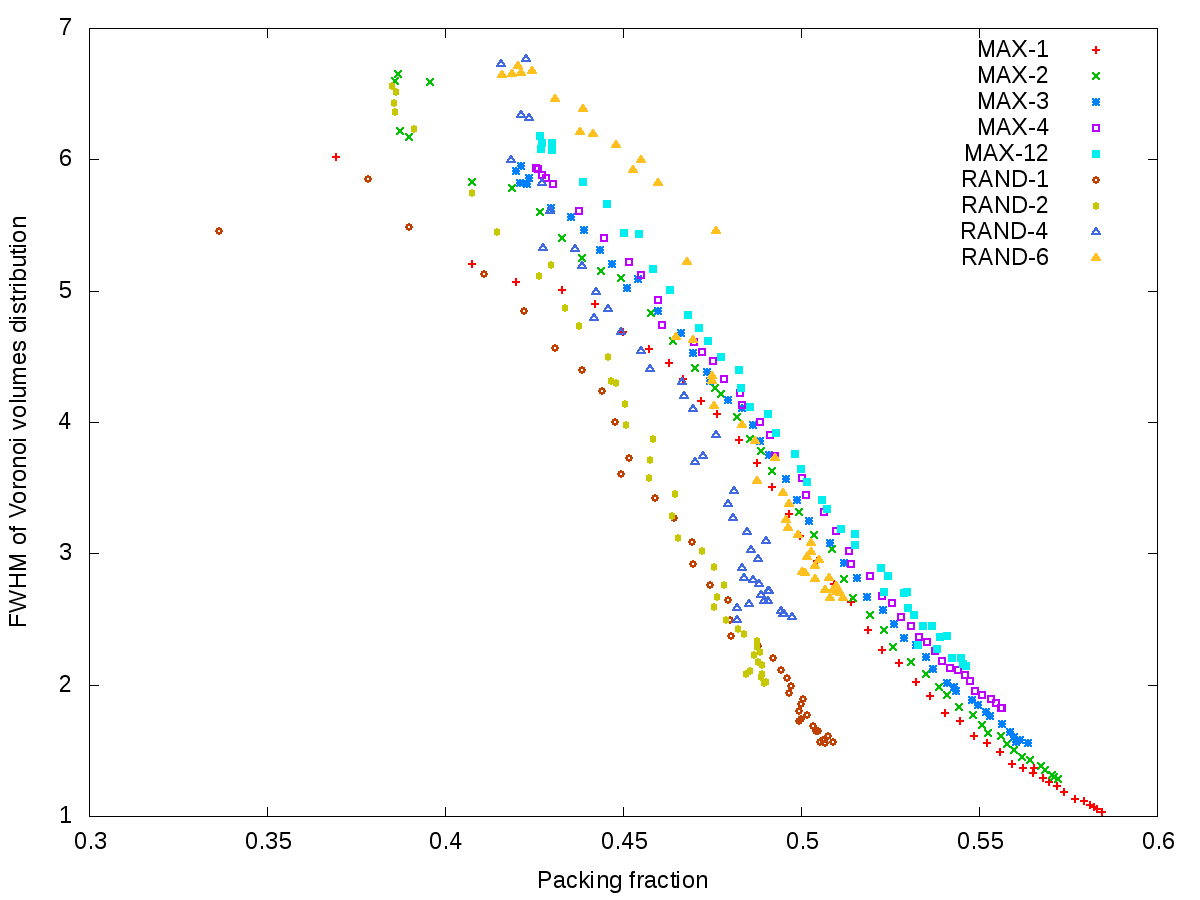}
\caption{Variations of the FWHM of the distributions of Voronoi cells volume with packing fraction. \label{fig:FWVor}}
\end{center}
\end{figure}

Motivated by Edward and Oakeshott's seminal work on statistical mechanics of granular systems \cite{EO89}, Aste and Di Matteo \cite{ADM08} have shown that, on a number of experimental as well as numerical random aggregates, the Voronoi cell volumes follow a so-called k-Gamma distribution law (based on Poisson distribution), which writes:
\begin{equation}
f(V) = \frac{k^k}{\Gamma(k)} \frac{(V-V_{min})^{k-1}}{(\bar V - V_{min})^k}\exp\left( -k\frac{V-V_{min}}{\bar V - V_{min}}\right)
\label{eq:kgamma}
\end{equation}
As can be seen in figure \ref{fig:distvo}, the distributions of the Voronoi cell volumes of the present aggregates are also well described by relation \ref{eq:kgamma}. However, when determining parameters by the least square method, the high sensitivity of $k$ to the precise value of $V_{min}$ did not allow a satisfying comparison of the dependency of $k$ with the packing fraction.
However, the FWHM of these distributions can be examined (figure \ref{fig:FWVor}). The variation of the FWHM appears linear over the whole packing fraction range and presents no discontinuity.

\begin{figure}[htbp]
\begin{center}
\includegraphics[width=0.8\textwidth]{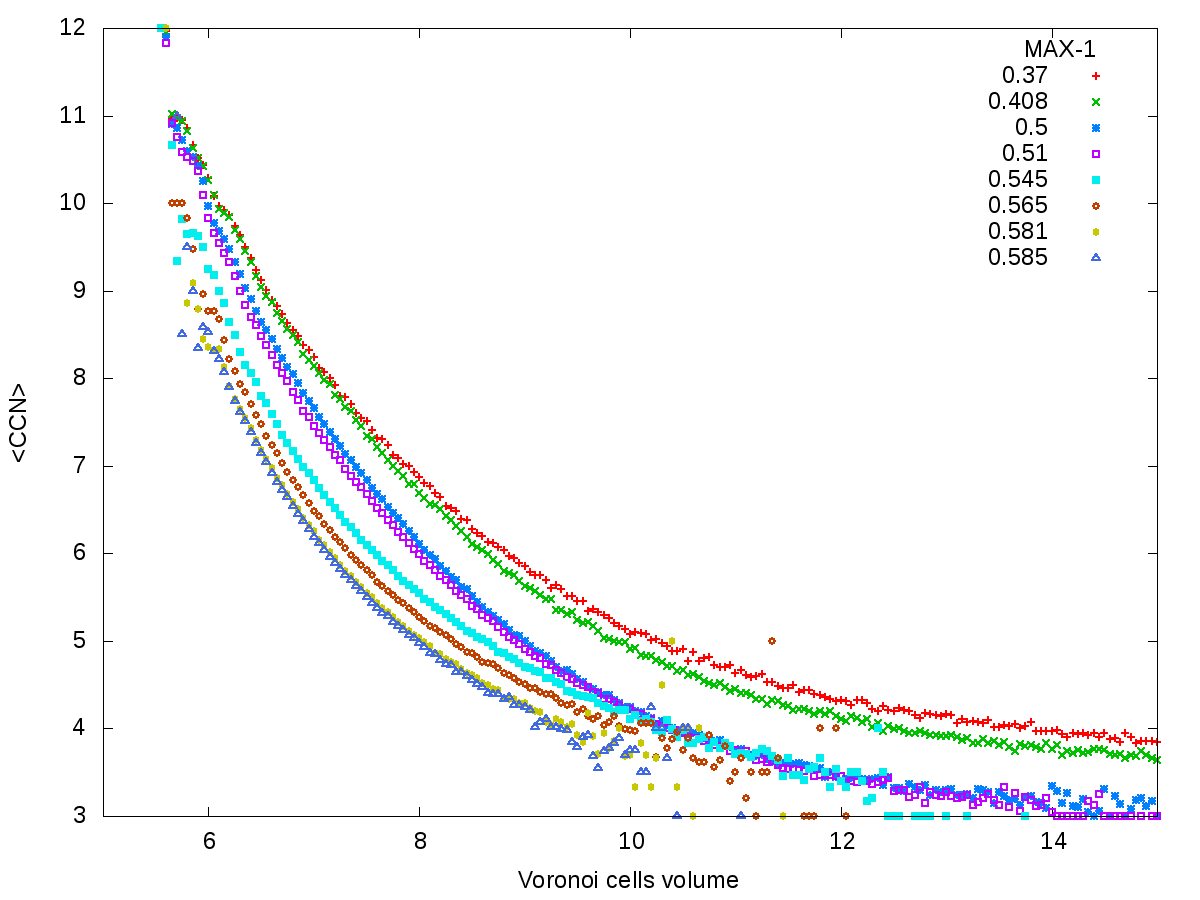}
\caption{Variations of the average contact coordination number of the spheres with their associated Voronoi cell volume. V is in $r_s^3$.\label{fig:bcooVoro}}
\end{center}
\end{figure}

The relation of the average contact coordination number of spheres with their associated Voronoi volume could be extracted (figure \ref{fig:bcooVoro})--the smaller the Voronoi cell volume (i.e. the more "crowded" the sphere environment), the higher the average contact coordination number. More quantitatively, this dependency has the form $\bar\eta \propto 1/V^{\zeta}$, where $\zeta$ roughly varies between 0.4 and 0.7. In the case of dyanmically built random aggregates, Wang et al. \cite{WSJM11} expect $\zeta = 1$.

\begin{figure}[htbp]
\begin{center}
\includegraphics[width=0.8\textwidth]{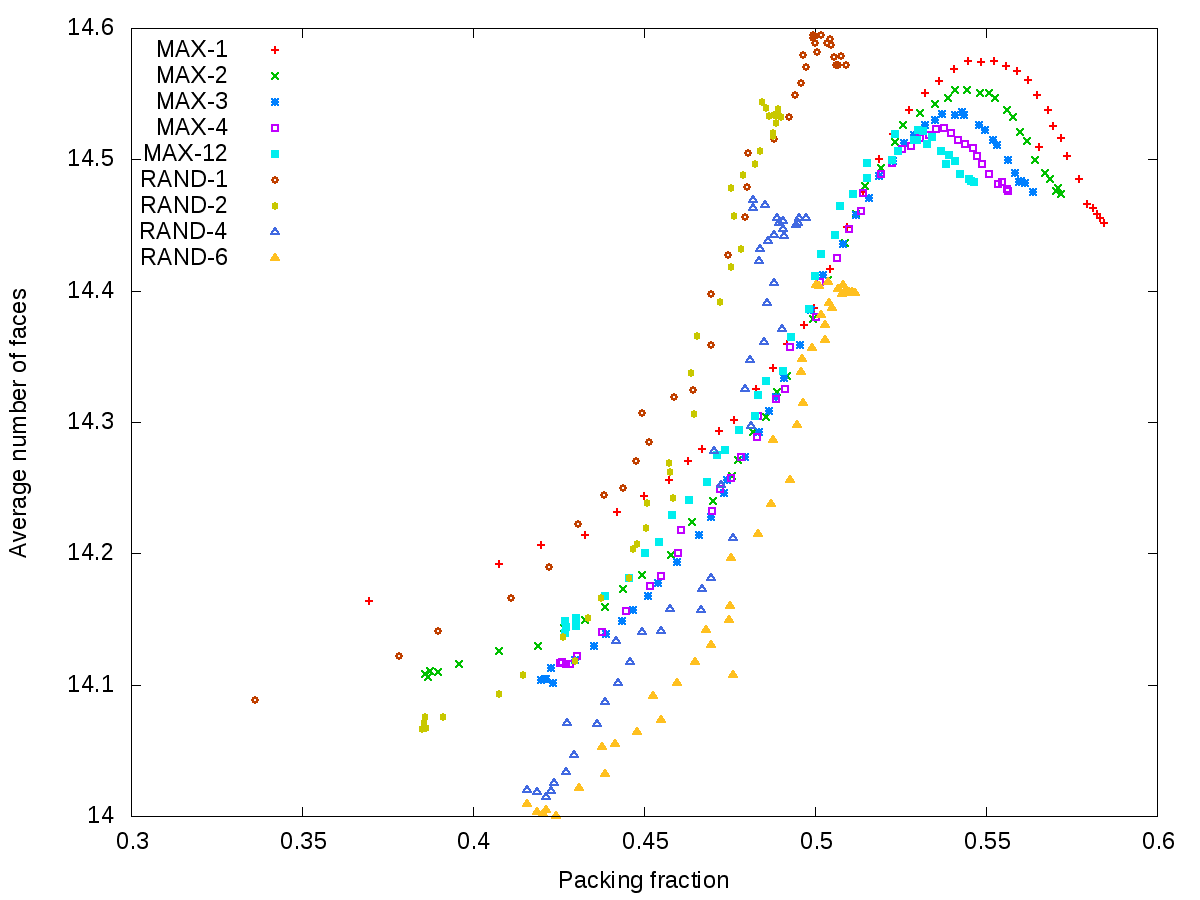}
\caption{Average number of faces of Voronoi polyhedra as a function of packing fraction.\label{fig:Faces}}
\end{center}
\end{figure}

The average number of faces $\bar F$ of the Voronoi cell i.e. the overall number of (contacting or non contacting) first neighbours of a given sphere is plotted in figure \ref{fig:Faces} as a function of packing fraction. $\bar F$ varies slightly between 14 and 14.6 and exhibits a maximum between $\gamma = 0.47$ and $\gamma = 0.54$ depending on the packing algorithm. In the case of aggregates generated using dynamic methods, the average number of faces of Voronoi cells decreases from 15.3 with $\gamma = 0.188$ down to 14.41 for $\gamma = 0.605$, according to Yang et al. \cite{YZY02}. The reason for this difference in behaviour between both families of aggregates is unclear.

\subsection{Pair distribution function}

Typical variations of the pair distribution function with packing fraction are presented in figure \ref{pdrtyp}. They are not mere homotethies, unlike the ones that are obtained by removing randomly spheres from a denser aggregate, which share the same PDF \cite{B77}.

\subsubsection{First neighbours}
\label{sec:fn}
\begin{figure}[htbp]
\begin{center}
\includegraphics[width=0.7\textwidth]{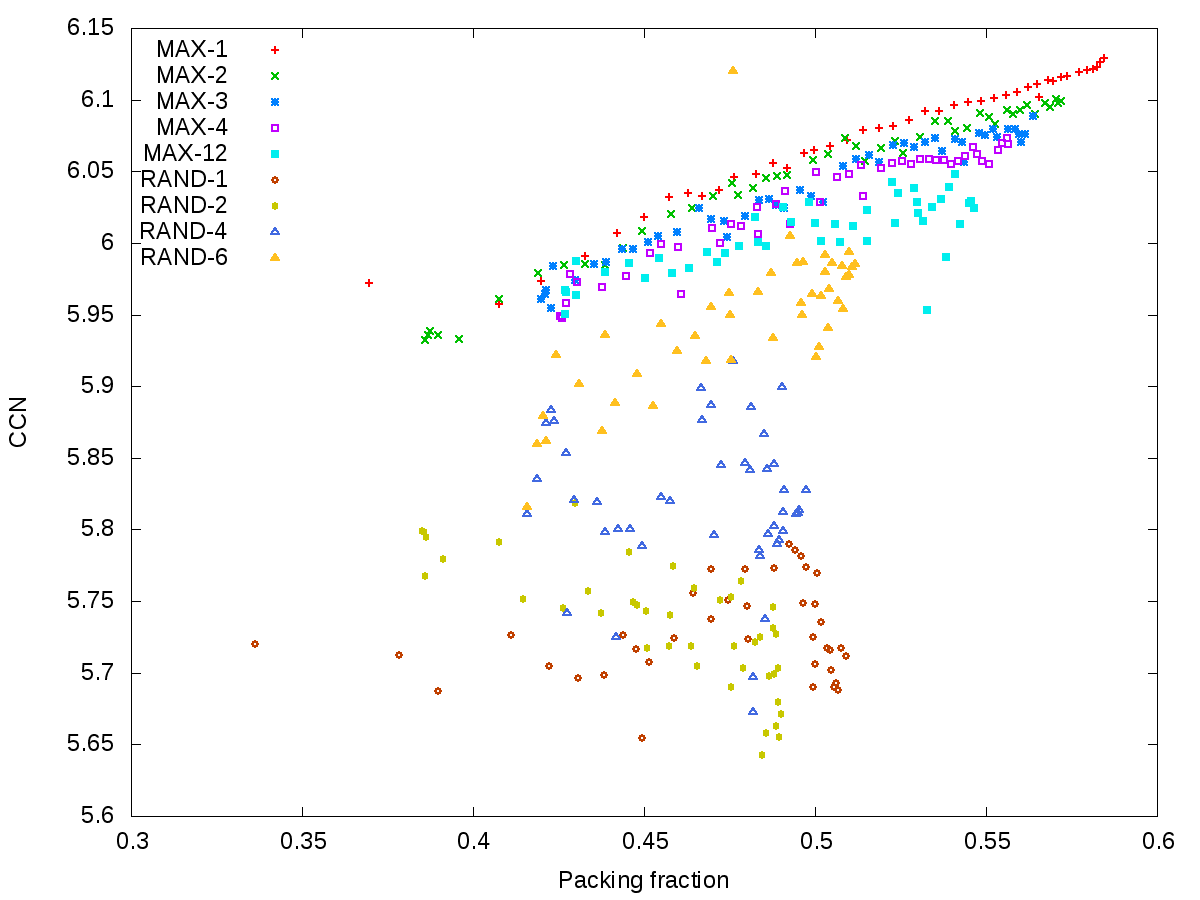}
\caption{CCN deduced from $P(r)$ as a function of packing fraction.\label{compapr}}
\end{center}
\end{figure}

\paragraph{Contact first neighbours}
From the intensity of the contacting spheres $\delta$ peak at $r=d$ in $P(r)$, it is possible to derive the average contact coordination number of a sphere $\bar\eta$ (relation \ref{deltapvan}), see figure \ref{compapr}. Overall, these results agree reasonably with the value of $\bar \eta$ derived above and show a very limited variation with packing fraction, which is classically observed for sequentially built random aggregates, as already mentioned above. However, it seems that the behaviour of the RAND family of aggregates differs significantly from the MAX family, but a more thorough examination of these aggregates properties is needed to determine the origin of such strong differences and will be the object of a forthcoming study.

\begin{figure}[htbp]
\begin{center}
\includegraphics[width=0.7\textwidth]{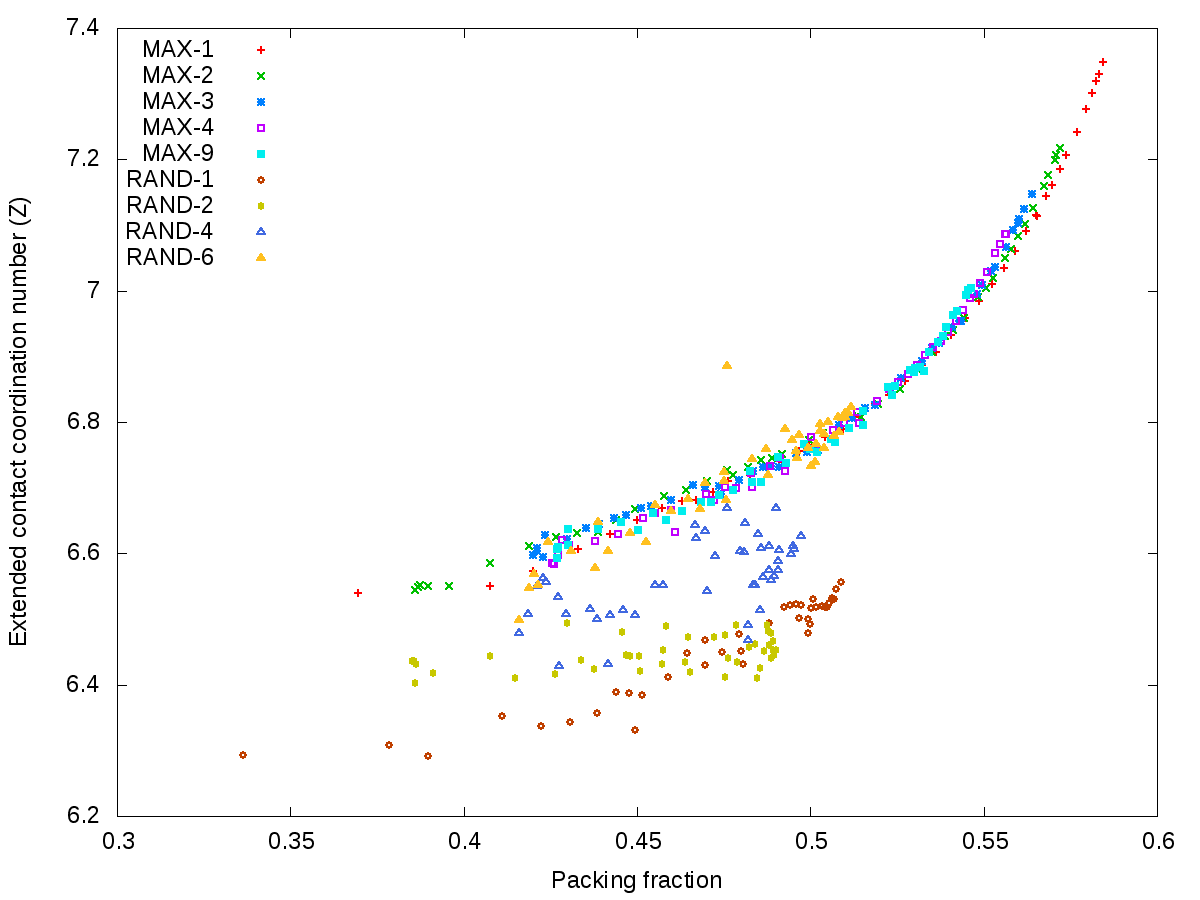}
\caption{Dependence of the extended contact coordination number $Z$ with packing fraction.\label{fig:extneig}}
\end{center}
\end{figure}

\paragraph{Extended first neighbours}
For pair distribution functions corresponding to different packing fractions, strong differences appear in $P(r)$ near the contact $\delta$ peak, which correspond to nearby first neighbours (figure \ref{pdrtyp}). 
The densest aggregates present the largest values of $P(r)$ in the interval $d<r<1.4d$, where $1.4d$ is the approximate position of the minimum separating the first neighbour peak from the second neighbour peak in $P(r)$. 
Adding the contribution of these extended first neighbours to the contacting neighbours gives the "total" number of first neighbours, $Z$, which writes:
\begin{equation}
 Z = \bar \eta + \rho\int_{d+\sigma}^{1.4d}P(r)4\pi r^2dr
\label{eq:eta2}
\end{equation}
$Z$ increases with the packing fraction and presents two growth regimes (figure \ref{fig:extneig}). The transition occurs between $\gamma=0.5$ and $0.55$. These extended first neighbours values can be reasonably well extrapolated to the maximum value of $\eta\approx8$ found by Bernall in the RCP. The fact that the present families of aggregates do not show this behaviour for $\bar\eta$ but rather for $Z$ might be due to the absence of medium to long range rearrangements due to the static nature of the building algorithms, which must have short range consequences and influence the optimal reorganisation of first neighbours, as already noted.

\subsubsection{Second and farther neighbours}
\label{sec:secneigh}
\begin{figure}[htbp]
\begin{center}
\subfigure[]{\includegraphics[width=0.4\textwidth]{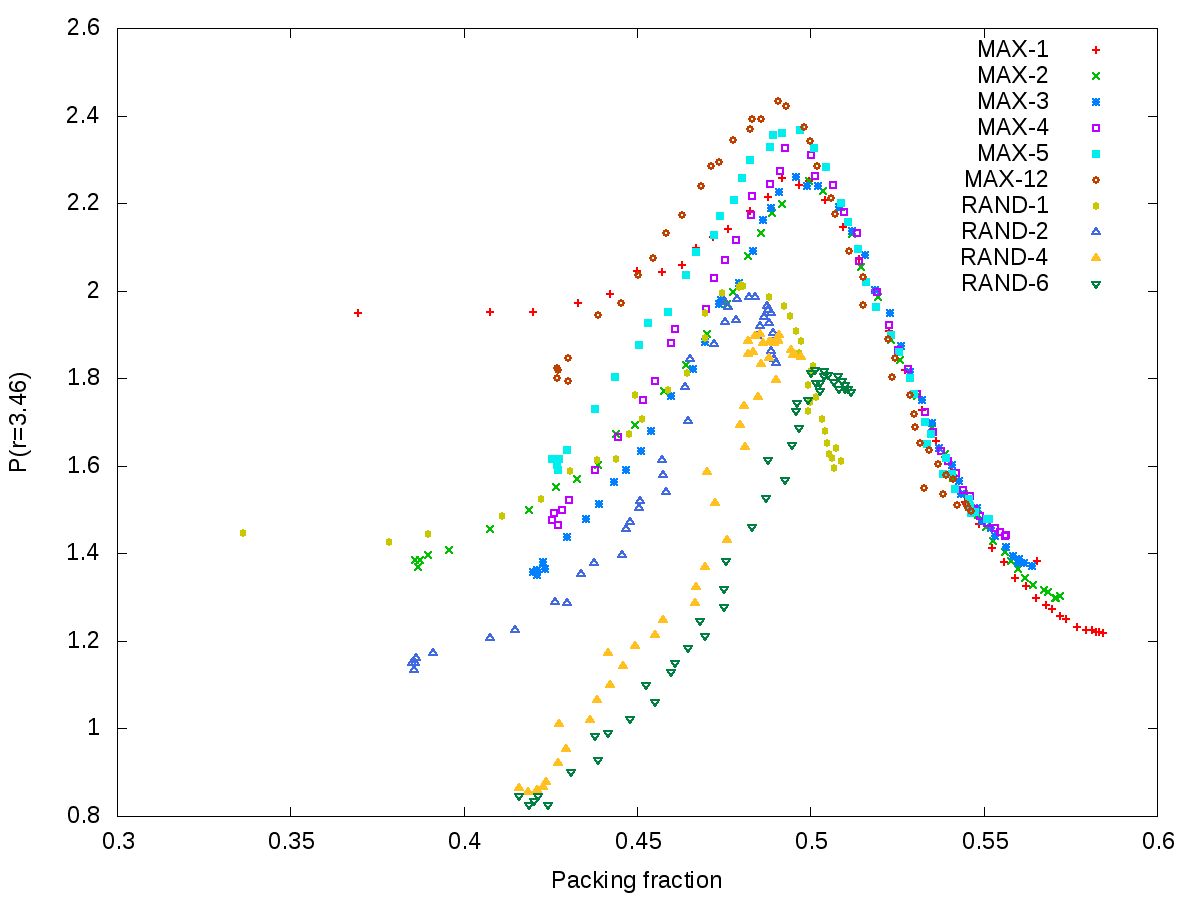}}
\subfigure[]{\includegraphics[width=0.4\textwidth]{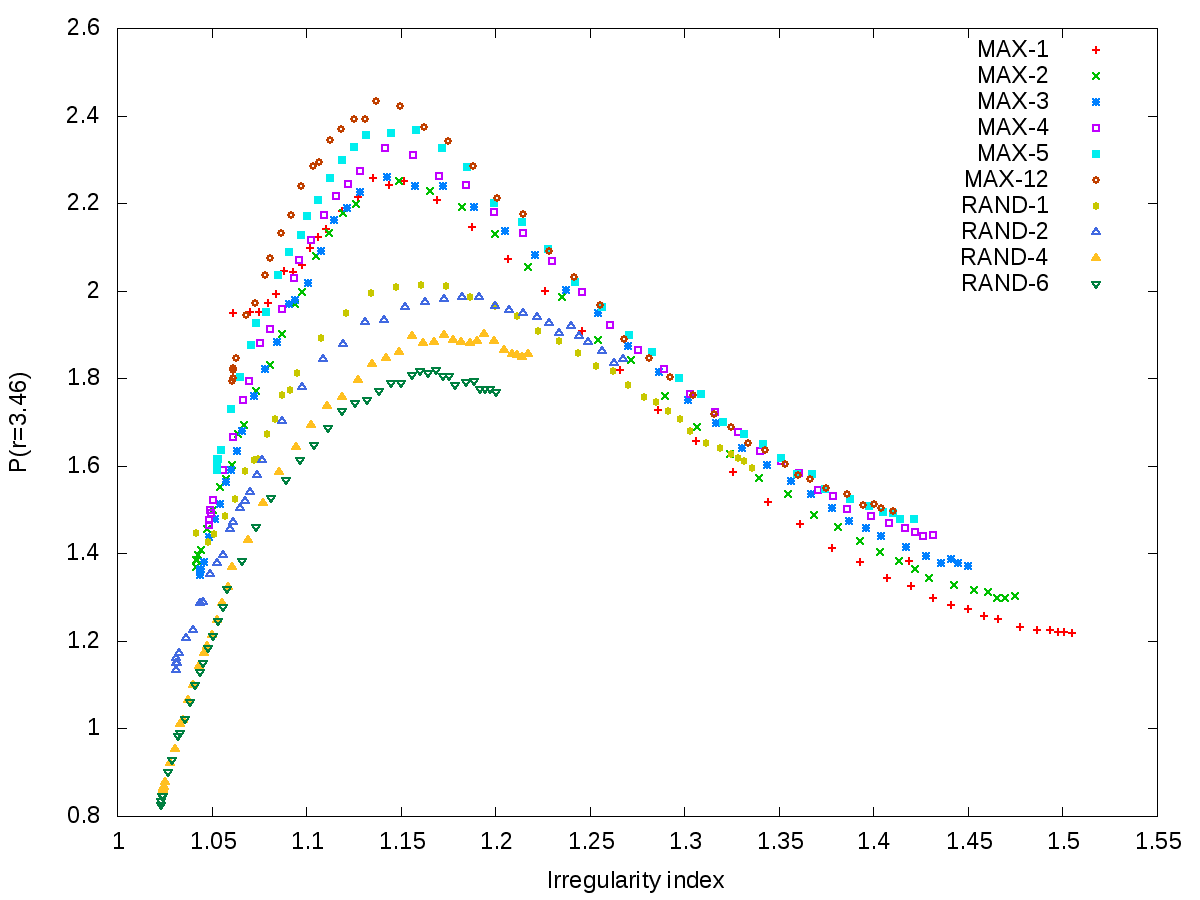}}
\subfigure[]{\includegraphics[width=0.4\textwidth]{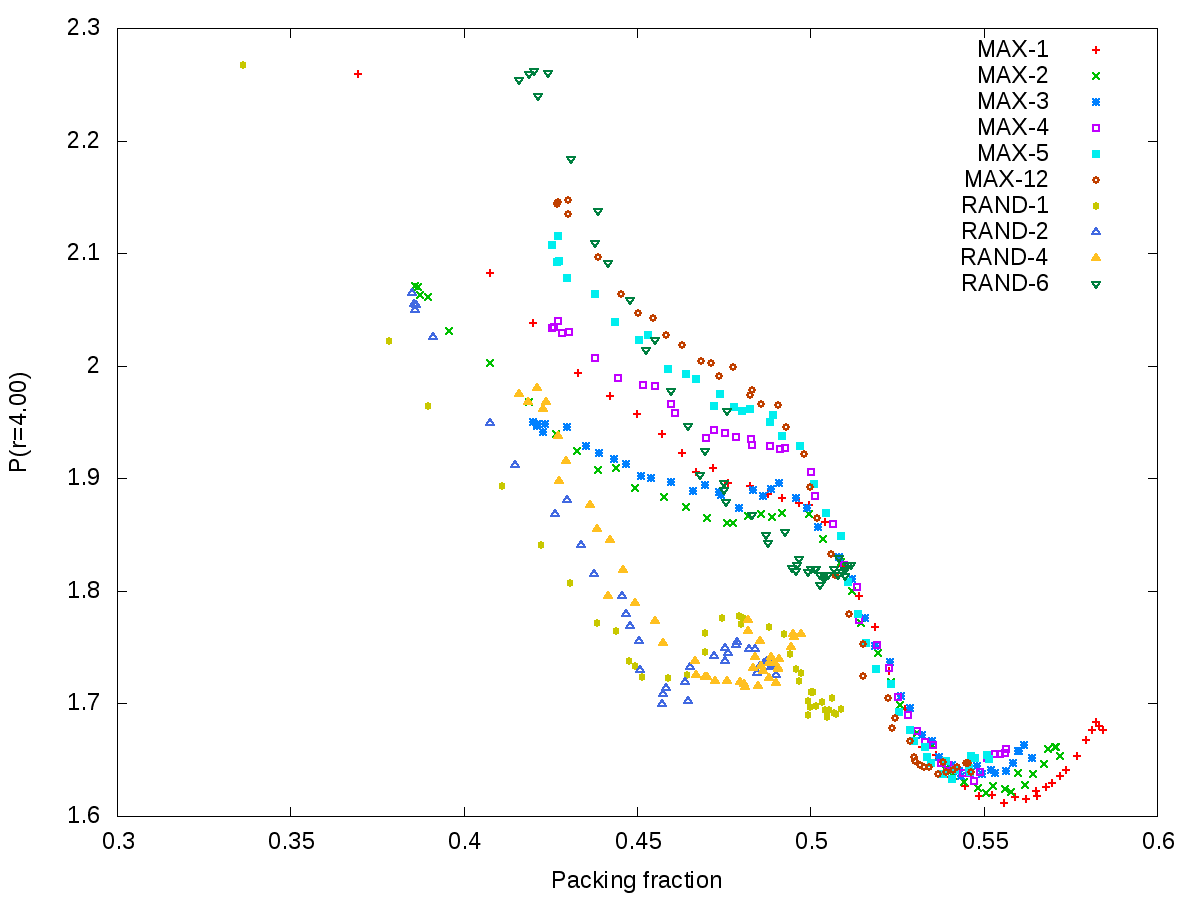}}
\caption{Variation of the intensity $P(r=\sqrt{3}d)$ as a function of packing fraction (a) and the irregularity index (b). Variation of the intensity of $P(r=2d)$ as a function of packing fraction (c).\label{fig:maxima}}
\end{center}
\end{figure}

\paragraph{Topological second neighbours maxima} Independently of the algorithm used, the second neighbours peak in $P(r)$ exhibits two sharp "sub-maxima" at $r=\sqrt{3}d$ and $r=2d$ whose origin is topological (figure \ref{pdrtyp}). The maximum at $r=\sqrt{3}d$  corresponds to the positioning triplet for which $\kappa$ is maximum ($\kappa=2$) (2-dimensional) and the maximum at $r=2d$ corresponds to the maximum second neighbours distance between three (aligned) contacting spheres (1 dimensional). Figure \ref{fig:maxima}.a and \ref{fig:maxima}.c show that the intensity of the triplet peak exhibits a maximum value of about 2.4 around a packing fraction of 0.5 while the peak at $2d$ decreases from about 2.3 when the packing fraction increases, with an inflexion at $\gamma \approx 0.5$.

Moreover, comparing figures \ref{fig:maxima}.a and \ref{fig:maxima}.b, which display the dependency of $P(r=\sqrt{3}d)$ respectively with $\gamma$ and $\bar\kappa$, shows that using $\bar\kappa$ rather than $\gamma$ produces much smoother curves, indicating that the average irregularity index is a better suited parameter to describe the aggregates structure than the packing fraction.

\subsubsection{$\delta$-peaks due to regular polytetrahedra}

The most appealing result on $P(r)$ is the contribution of regular polytetrahedra with edge length $d$ that could be detected thanks to the precision of these calculations coming from the large number of spheres in the aggregates. This contribution increases as the packing fraction decreases. 
The pair distribution function \ref{pdrnum} can then be rewritten as:
\begin{equation}
P(r) = \frac{\bar\eta}{4\pi \rho d^2} \delta (r-d) + \sum_{p,d_p>2} \frac{\bar\eta_p}{4\pi\rho d_p^2} \delta(r-d_p) + P_c(r)
\label{pdrpoly}
\end{equation}
where
\begin{itemize}
\item the first $\delta$ peak at $r = d$ describes the unavoidable contribution of contacting spheres,
\item $d_p$ is the series of the discrete distances between two non contacting sphere (centers) belonging to the same regular polytetrahedron. The peaks associated with these polytetrahedral pairs are represented by $\delta(r-d_p)$ distributions because there is a finite number of such pairs in a length interval of null width (in analogy with the case of contacting pairs of spheres) .
\item $\bar\eta_p$ is the average number of sphere centers at a distance $d_p$ from a given sphere center
\item The third term $P_c(r)$ represents the contribution of all other pairs of sphere centers and does not involve any $\delta$ singularity.
\end{itemize}

As the number of vertices $s$ of the regular polytetrahedra increases new distances appear in the distance series involved in equation \ref{pdrpoly}.

\begin{figure}[htbp]
\begin{center}
\subfigure[]{\includegraphics[width=0.4\textwidth]{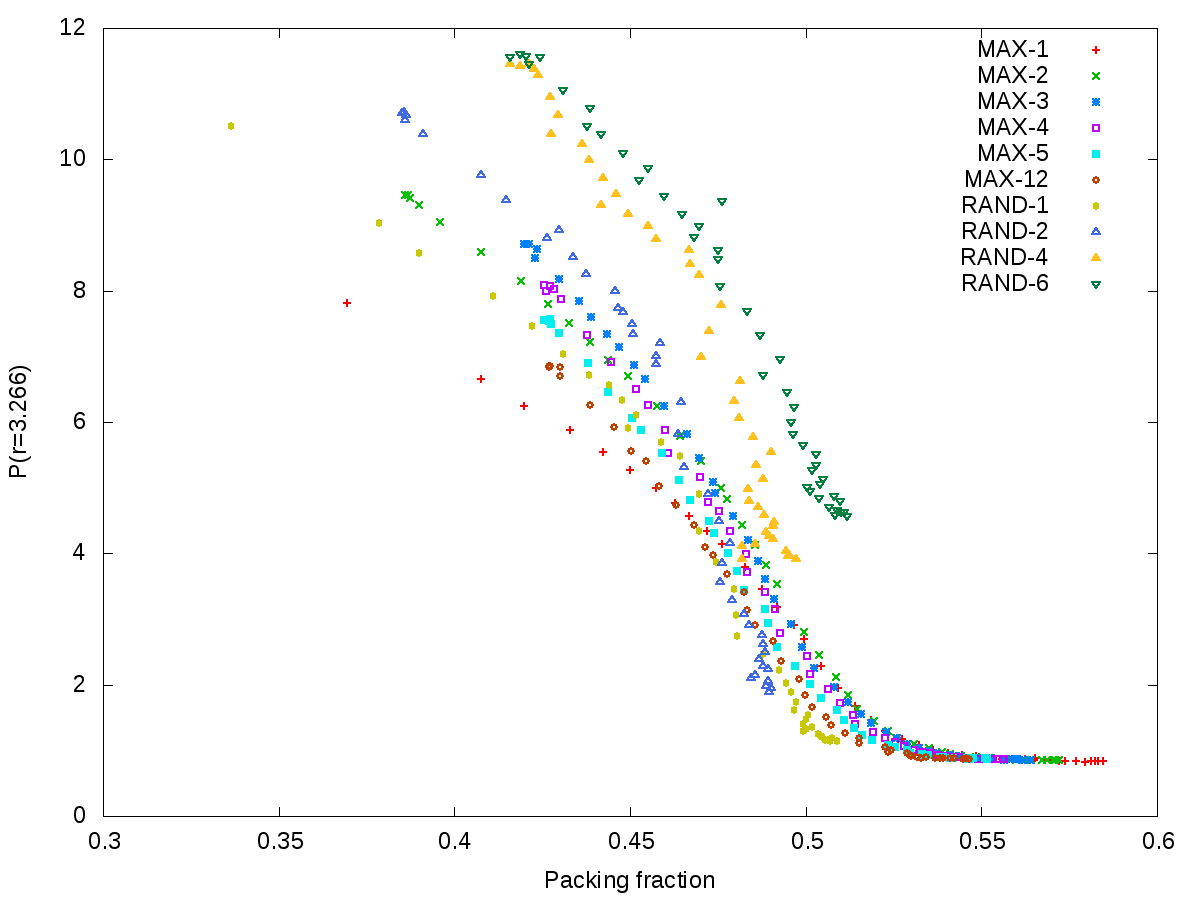}}
\subfigure[]{\includegraphics[width=0.4\textwidth]{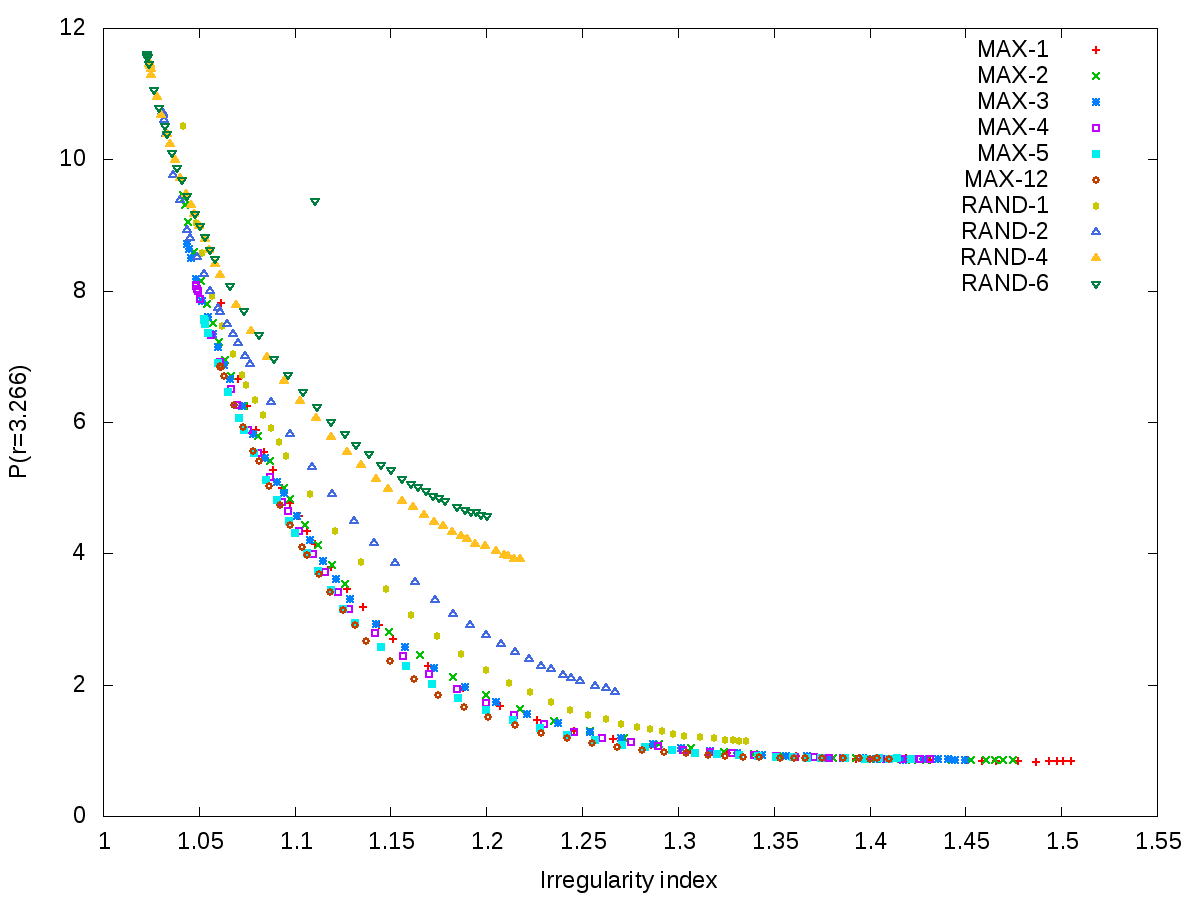}}
\caption{Variation of the intensity $P(r=d\sqrt{8/3})$ as a function of a) packing fraction b) irregularity index.\label{fig:p326})}
\end{center}
\end{figure}

The first one to appear, at $d_1 = d\sqrt{8/3}$, corresponds to the 5 vertices bi-pyramid. Figure \ref{fig:p326} shows that the (numerical) intensity of this $\delta$ peak increases very rapidly to large numerical values (larger than 10) with decreasing packing fraction and can be detected from $\gamma\approx0.52$ or $\kappa \approx 1.3$. As can be seen by comparing figure \ref{fig:p326}.a and \ref{fig:p326}.b, this intensity has a smoother dependence on $\kappa$ than on $\gamma$, suggesting once more that $\kappa$ is a better structural characteristic of the aggregate.

\begin{figure}[htbp]
\begin{center}
\includegraphics[width=0.7\textwidth]{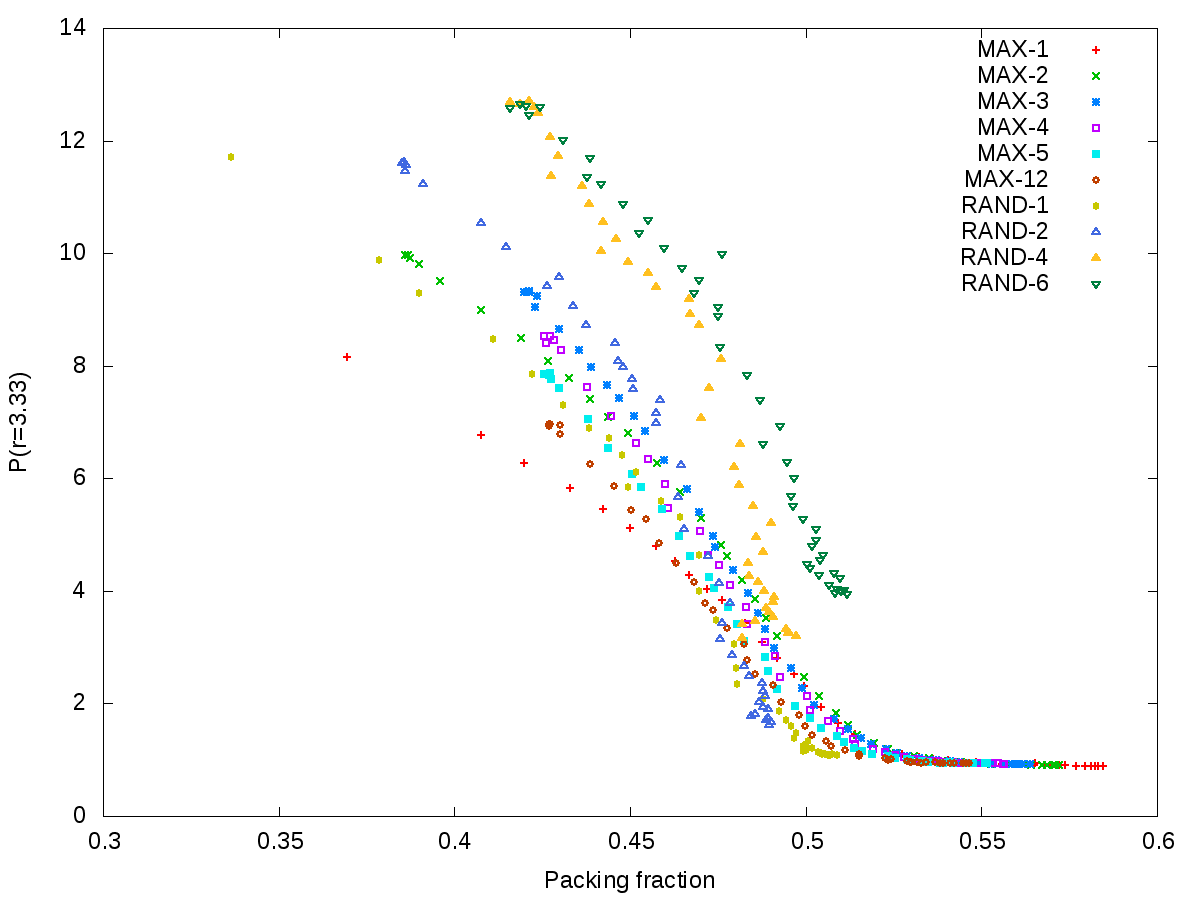}
\caption{Variation of the intensity $P(r=5d/3)$ as a function of packing fraction.\label{fig:p333}}
\end{center}
\end{figure}

The second $\delta$ peak to appear at $d_2= 5d/3$ corresponds to the unique regular polyhedron with 6 vertices. Its intensity behaves in the same way as that of the peak at $r=d_1$ vs. $\gamma$ and $\kappa$, but appears for slightly lower values of these two parameters as it grows from the 5-vertices polyhedron (figure \ref{fig:p333}).

Beyond $s = 6$, each vertex addition can give rise to several polytetrahedral isomers with the same total number of vertices. Furthermore, for a given isomer, the last added vertex also gives rise to several new pair distances $d_p$. As a result, the number of terms of the distance series increases rapidly with $s$ and more $\delta$ peaks appear at lower $\gamma$ values due to the progressive growth of the polyhedra with higher vertices number.

At low packing fraction (below $\gamma=0.52$), the proportion and size of the regular polyhedra first increase but they remain embedded in a more disordered "matrix" made from irregular (sphere centers) tetrahedra... until, intuitively, the system reaches a percolation threshold where an infinite (i.e. running over the whole aggregate) regular polytetrahedron is formed which coexists white finite regular polytetrahedra and irregular tetrahedra as well. However we could not determine the packing fraction corresponding to this threshold. 
The underlying mathematics of this description is still to be deepened.

The loose packed aggregates studied here have a composite structure made from a polyhedral phase (which could be characterized by the appearance of discrete $\delta$ peaks in $P(r)$) coexisting with a more disordered phase made from irregular tetrahedra (of sphere centers). Conversely, the dense packed aggregates involve a unique disordered phase made from irregular tetrahedra.

\begin{figure}[htbp]
\begin{center}
\includegraphics[width=0.7\textwidth]{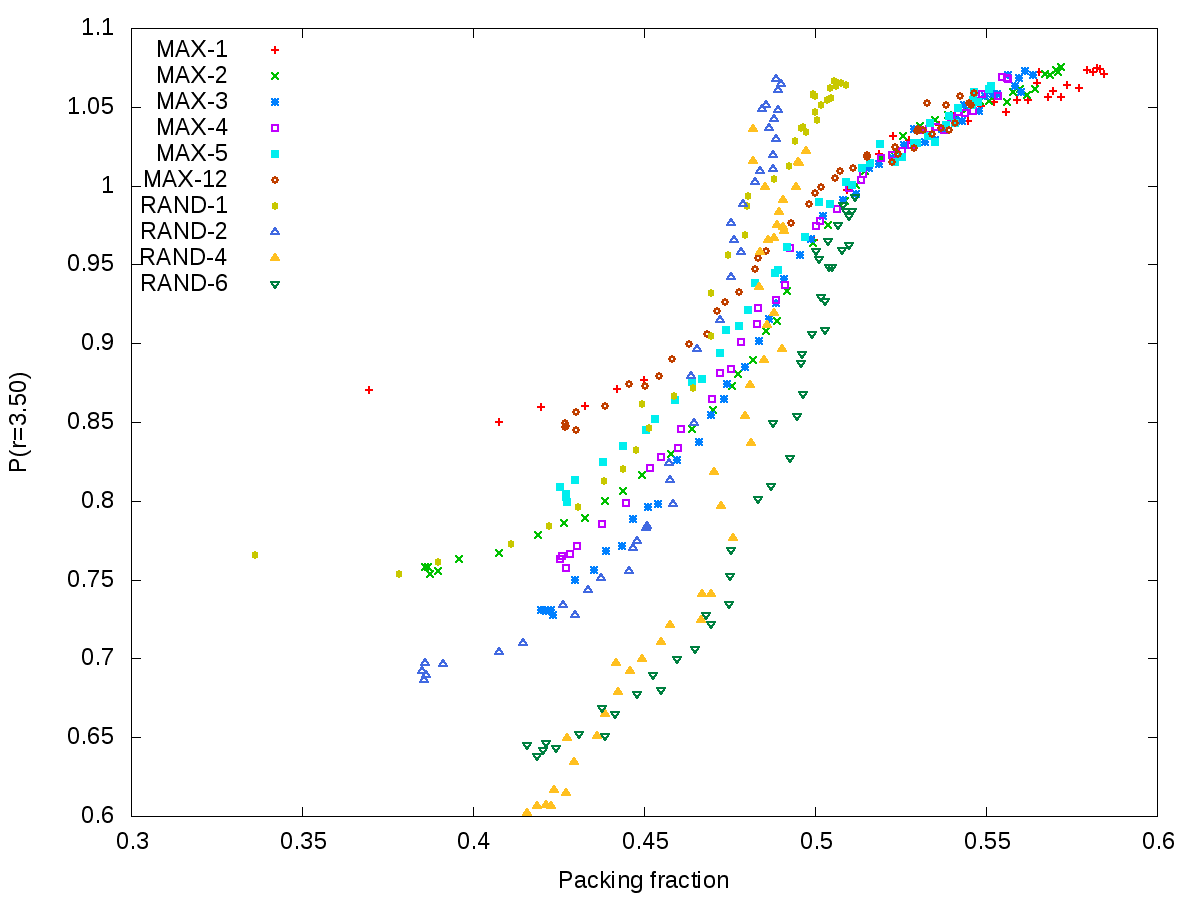}
\caption{Variation of the intensity $P(r=1.75d)$ with packing fraction. This point do not correspond to a $\delta$ peak and is representative of the continuous part of the total PDF.\label{fig:p350}}
\end{center}
\end{figure}

Finally, the continuous (free from $\delta$ peaks) contribution to the PDF, $P_c(r)$, is illustrated in figure \ref{fig:p350}, by the intensity curve corresponding to an arbitrary value of $P(r=1.75d)$, which lies in the second neighbours range. The intensity of this arbitrary point increases with $\gamma$ and $\kappa$, and displays a lower growth rate beyond $\gamma \approx 0.51$. The behaviour of the neighbours beyond the second ones is analogous but damped by the general decrease of $P(r)$ oscillations at large $r$, which goes to 1 at large $r$.

\subsection{Structure factor}
\subsubsection{Density fluctuations and small Q behaviour of S(Q)}
\begin{figure}[htbp]
\begin{center}
\includegraphics[width=0.8\textwidth]{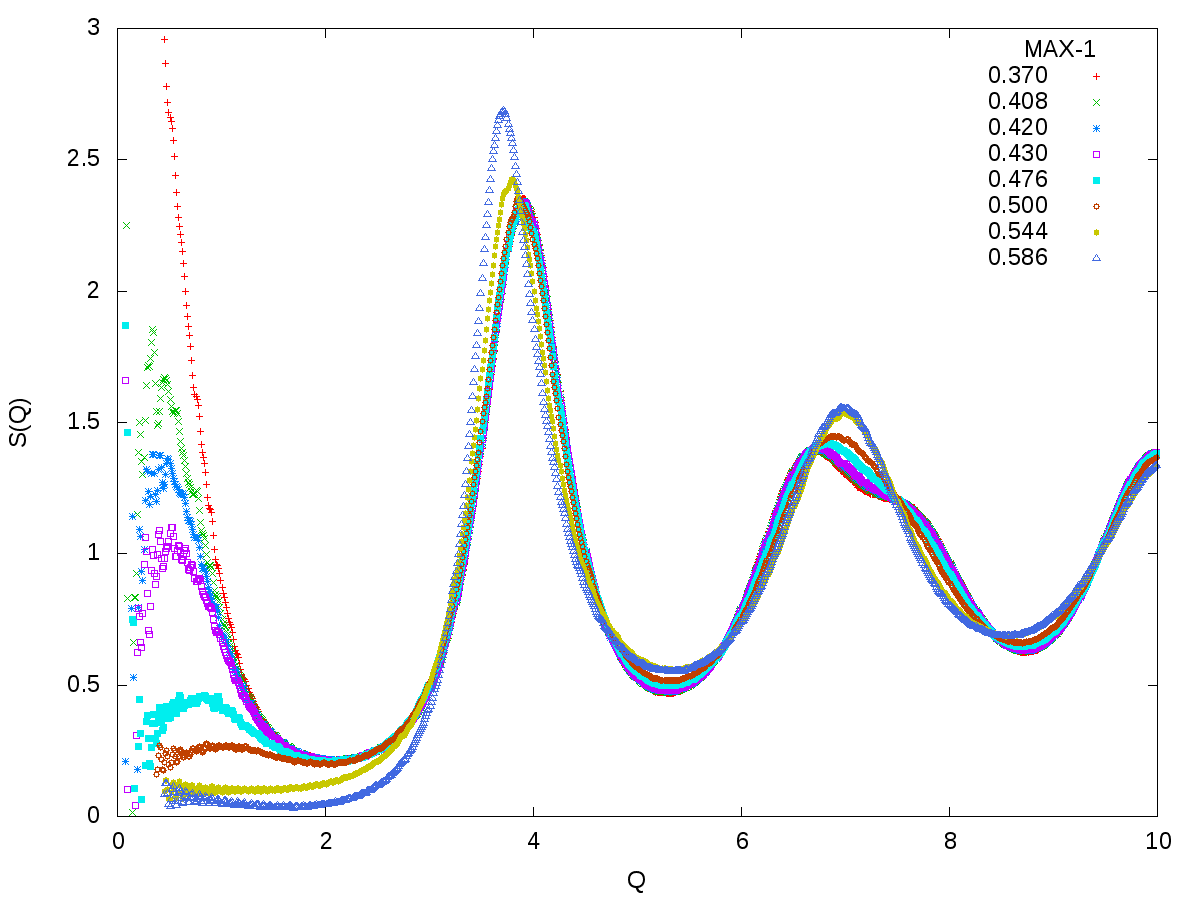}
\caption{Structure factor for various packing fraction. $S(0)$ values were taken from fluctuations calculations (relation \ref{eq:scinap4}). $Q$ is in $r_s^{-1}$ unit.
\label{sq_compa}}
\end{center}
\end{figure}

The large number of spheres of the aggregates studied here allows the study of the behaviour of the structure factor at small $Q$ values.
Figure \ref{sq_compa} compares the structure factor obtained for various aggregates. 
Thanks to the small angle scattering correction presented in section \ref{sec:cordif}, it is possible to remove the aggregate shape contribution down to $Qr_s\approx 0.4$. 

As $Q$ goes to 0, the $S(Q)$ corresponding to different packing fractions extrapolates well to the values of $S(0)$ calculated from the number density fluctuations (see section \ref{sec:sinap4}), with--for the highest packing fractions (i.e. $\gamma>0.5$)--a horizontal tangent at $Q=0$, according to the relation given by Donev et al. \cite{DST05}:
\begin{equation}
\lim\limits_{Q \to 0} S(Q) = S(0) + CQ^2
\label{eq:smallq}
\end{equation}
where $C$ is a constant.

In the interval $0<Q<Q_1$, where $Q_1$ is the position of the structure-factor first-pea--i.e. $Q_1 \in [3.65, 4]$--, the present calculations show that $S(Q)$ exhibit a minimum around $Qr_s = 1.7$ whose depth increases as the packing fraction decreases. 
This minimum seems to have been experimentally observed in structure factors of some liquid metals, whose structure is well represented by disordered sphere packings \cite{RE75, W80}.

For packing fractions below 0.5, an interference peak develops at small $Q=Q_\phi$ values, corresponding to distances in real space of about $d_\phi = 2\pi/Q_\phi$. 
Remembering that low density aggregates have a composite structure made of polyhedra embedded in a more disordered matrix, this peak is likely due to interferences between these polyhedra separated by an average distance $d_\phi$ (see \cite{G94}). Its intensity increases when the packing fraction decreases.

\subsubsection{First peak of the structure factor}

\begin{figure}[htbp]
\begin{center}
\includegraphics[width=0.7\textwidth]{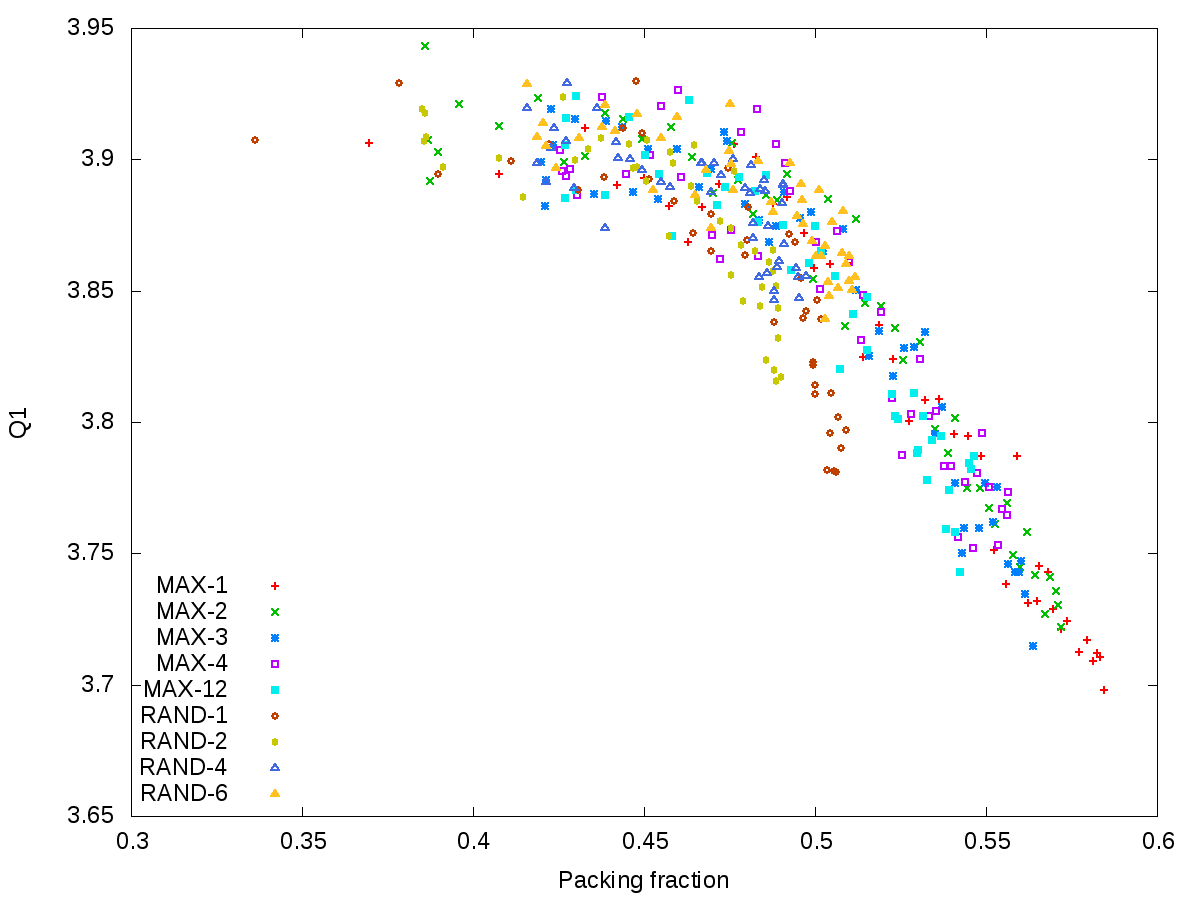}
\caption{Variations of the position of the first peak of the structure factor, $Q_1$, with packing fraction. $Q$ is in $r_s^{-1}$ unit.
    \label{fpicpos}}
\end{center}
\end{figure}

\begin{figure}[htbp]
\begin{center}
\includegraphics[width=0.7\textwidth]{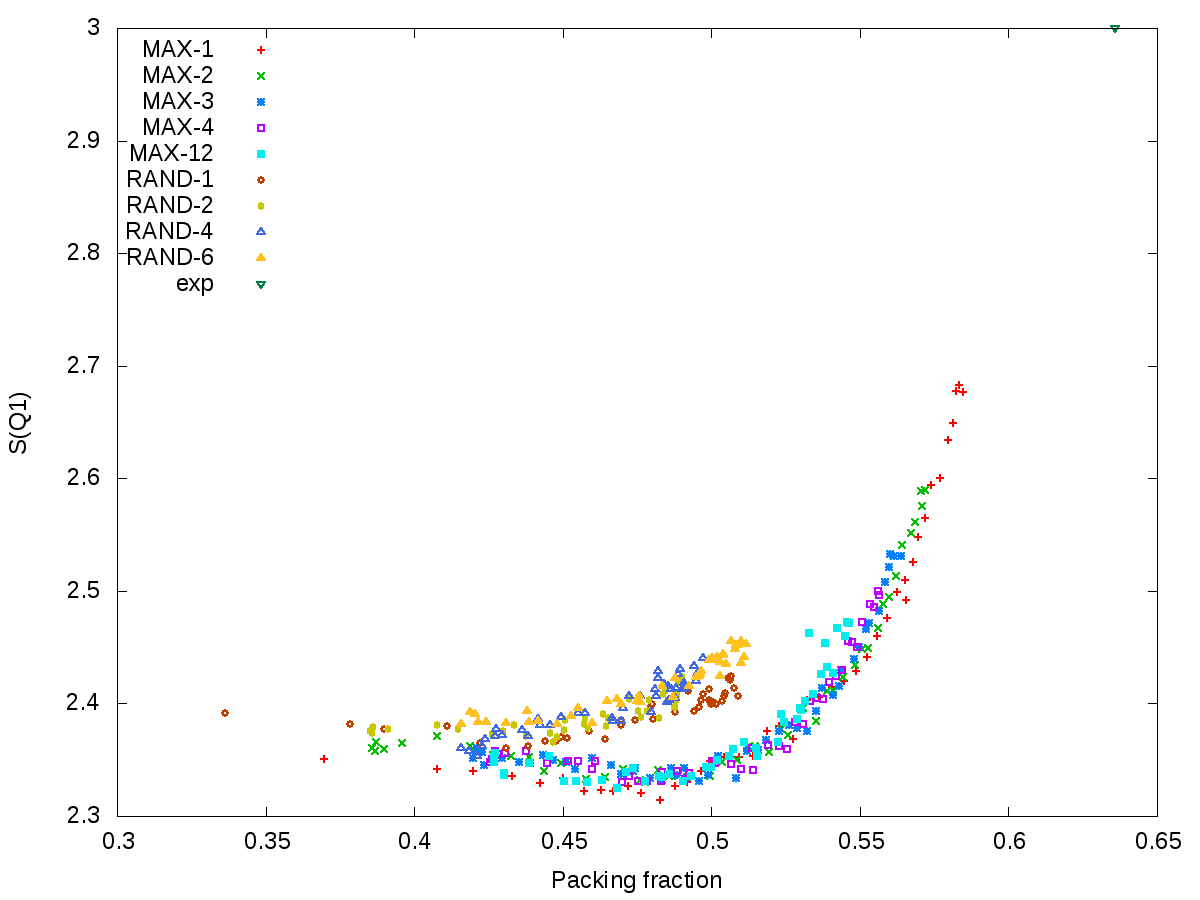}
\caption{
    Variations of the intensity of the first peak of the structure factor, $S(Q_1)$, with packing fraction.
    \label{fpicmax}}
\end{center}
\end{figure}

The position $Q_1$ and intensity $S(Q_1)$ of the first peak of the structure factor vary with the packing fraction of the aggregates (figures \ref{fpicpos} and \ref{fpicmax}). 
$Q_1$ presents a plateau for low packing fraction, up to $\gamma\approx0.48$ and then decreases linearly with $\gamma$. It extrapolates to $Q_1r_s = 3.59$ for the RCP. $S(Q_1)$ decreases slightly when the packing fraction increases up to $\gamma \approx 0.48$ and then increases with packing fraction and roughly extrapolates (linearly) to 3.1 for the RCP. This last value is in agreement with experimental values obtained on liquid metals \cite{W80,K76}.

\subsubsection{Second peak of the structure factor}

The shape of the second peak of the structure factor around $Qr_s \approx 7$ evolves continuously with packing fraction (figure \ref{sq_compa}). For packing fractions higher than about 0.48 to 0.52 (depending on the algorithm used), a single peak is observed. When the packing fraction decreases below this limit, a shoulder first appears and progressively transforms into a new peak. This is likely due to the progressive formation of polyhedra which will be further assessed by the high $Q$ behaviour of $S(Q)$. This shoulder might have been observed on some pure liquid metals \cite{W80}.

\subsubsection{Large Q behaviour}
The large $Q$ behaviour of the structure factor $S(Q)$, which is derived from the pair distribution function by the Fourier transformation \ref{eq:scinap3}, deserves special attention since it only depends on the $\delta$ peaks of $P(r)$ through relation:
\begin{equation}
\lim\limits_{Q \to \infty} S(Q) = 1 + \bar\eta \frac{\sin(Qd)}{Qd} + \sum_{p,d_p>2} \bar\eta_p \frac{\sin (Qd_p)}{Qd_p}
\label{eq:fourlarG}
\end{equation}

For dense aggregates, which are (almost) free from regular polytetrahedra, $S(Q)$ behaves as predicted by relation \ref{eq:scinap5} since the major Dirac peak in $P(r)$ is the peak of contacting neighbours at $r = d$. One therefore observes $S(Q)$ damped oscillations with $2\pi/d$ periodicity at large $Q$.

For low density aggregates, with $\gamma < 0.47-0.52$ (depending on the algorithm), i.e. when the contribution of regular polytetrahedron increases with decreasing packing fraction, the $\delta$ peaks observed in $P(r)$ have a weak intensity and do not change significantly the period of $S(Q)$ oscillations but only their detailed shape. Therefore it is very difficult to detect the formation of the first polytetrahedra by observing the asymptotic behaviour of $S(Q)$.

\begin{figure}[htbp]
\begin{center}
\includegraphics[width=0.8\textwidth]{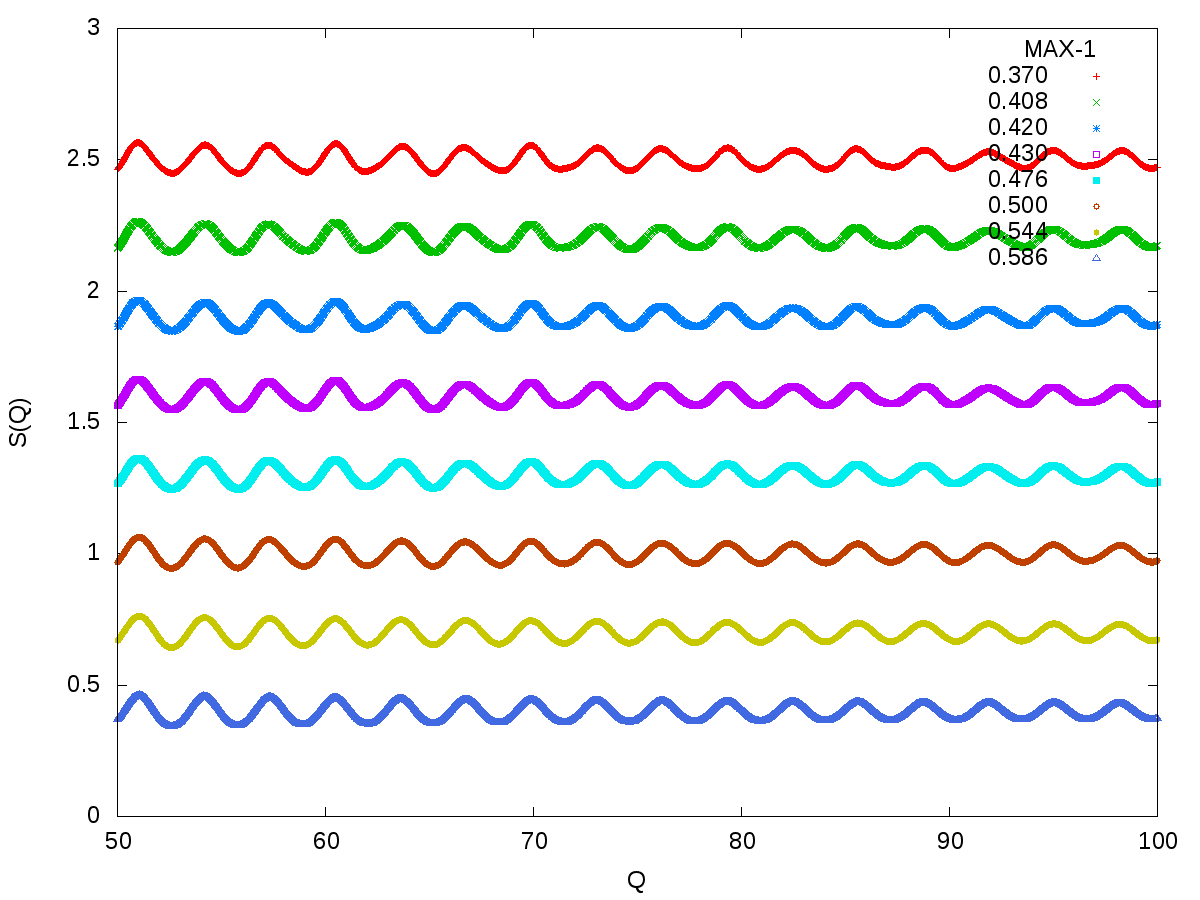}
\caption{Asymptotic behaviour of $S(Q)$--curves have been arbitrarily shifted vertically from their mean value 1. $Q$ is in $r_s^{-1}$ unit.\label{fig:Sasymp}}
\end{center}
\end{figure}

\subsubsection{Comparison with structure factor of liquid metals}
These structure factor calculations with increased accuracy showed their physical interest for an improved modelling of the experimental structure factor of liquid or amorphous metals and alloys in the dense packing regime. Three main results are worth mentioning:
\begin{itemize}
  \item The increased size of the spheres aggregates studied here allowed the determination of $S(Q)$ down to $Qr_s \approx 0.4$ and proved the existence of a weak "pre-minimum" around $Qr_s \approx 1.7$, i.e. before the structure factor first peak. That could explain old experimental results obtained by Reiter et al. \cite{RRS76} on the overall structure factor $S_{NN}(Q)$ of liquid Li - Ag alloys.
  \item The position $Q_1$ and intensity $S(Q_1)$ of the first peak of the structure factor were shown to vary slowly with the packing fraction and compare well with accurate experimental results on liquid metals compiled by Waseda \cite{W80} (especially near their melting point i.e. when atomic vibrations do not damp too strongly the structure factor).
  \item Finally, the existence of a shoulder on the second peak of $S(Q)$ could have been observed on amorphous Ni (\cite{I73}).
\end{itemize}

\section{Conclusion}

A wide class of random packings of sticky hard spheres built sequentially have been studied over the packing fraction range 0.329 to 0.586. However, packing fractions larger than 0.59 could not be reached by these static algorithms which can be understood since they optimize the sphere positioning up to the second neighbour distances only. More than 300 aggregates containing 10$^6$ spheres each were generated. They first allowed a study of local fluctuations. The sphere number density fluctuations, independently of the sphere packing algorithms used, were shown to follow a power law in $\gamma^{-3}$ and could be extrapolated to a very small value (less than $10^{-2}$) for the RCP network.

On the other hand, the high accuracy reached in the calculations allowed a careful study of the structural characteristics of the aggregates (coordination number, pair distribution function and structure factor). It turns out that the irregularity index of the positioning tetrahedra seems to be better suited than the packing fraction to classify some structural characteristics of the aggregates. Whatever the algorithm used, a transition from a low to a high density regime could be observed in the interval $0.48<\gamma<0.52$. It could be attributed to the formation and growth of regular polytetrahedra as the packing fraction decreases, which produces $\delta$ singularities in $P(r)$ and changes in $S(Q)$ (small $Q$ peak, shoulder on the second peak of $S(Q)$ and asymptotic behaviour). The low density aggregates therefore have a composite structure, made of regular polytetrahedra embedded in a more disordered matrix, while the high density aggregates are single phased. 

\bibliography{pap.bib}

\end{document}